\documentstyle[12pt]{article}
\def\footnoterule{\kern-3pt \hrule width \hsize \kern2.6pt}
\newcommand{\beq}{\begin{equation}}
\newcommand{\eeq}{\end{equation}}
\newcommand{\ba}{\begin{array}}
\newcommand{\ea}{\end{array}}
\newcommand{\beqa}{\begin{eqnarray}}
\newcommand{\eeqa}{\end{eqnarray}}
\newcommand{\bd}[1]{ \mbox{\boldmath $#1$}  }

\newcommand{\tresj}[6]{ \left( \begin{array}{ccc}
                               #1 & #2 & #3 \\
                               #4 & #5 & #6
                             \end{array}
                        \right) }

\newcounter{pepe}
{\end{eqnarray}%
\setcounter{equation}{\arabic{pepe}}%
}

\newcommand{\negra}[1]{\mbox{\boldmath$#1$}}

\newcommand{\nsigma}{\negra{\sigma}}

\begin{document}
\begin{titlepage}
\mbox{}

\noindent
\begin{center}
{\large {\bf
Parity violation in  quasielastic electron scattering \hfill \\[1.0ex]
from closed-shell nuclei$^*$}}

\vspace*{0.25in}

\noindent

J.E. Amaro$^{1,3}$, J.A. Caballero$^2$, T.W. Donnelly$^1$,
A.M. Lallena$^3$, E. Moya de Guerra$^2$ \hfill\\
and J.M. Ud\'\i as$^4$ \hfill\\[3mm]

{\it
$^1$ Center for Theoretical Physics, Laboratory for Nuclear Science,
and Dept. of Physics, \hfill \\
Massachusetts Institute of Technology,
Cambridge, MA 02139, USA \hfill\\[2mm]

$^2$ Instituto de Estructura de la Materia, Consejo Superior de
Investigaciones Cient\'{\i}ficas, \hfill\\
Serrano 123, Madrid 28006, Spain \hfill\\[2mm]

$^3$ Departamento de F\'{\i}sica Moderna, Universidad de Granada, \hfill\\
E-18071 Granada, Spain \hfill\\[2mm]

$^4$ National Laboratory for Nuclear and High-Energy Physics,
Section K (NIKHEF--K), \hfill \\
P.O. Box 41882, NL--1009 DB Amsterdam, The Netherlands}
\\
{~~~} \\
{\rm MIT-CTP-2472.  {~~~} nucl-th/9510006 {~~~} October 1995.} \\
{\rm Submitted to {\it Nuclear Physics A}}

\end{center}

\vspace{2\fill}

\begin{quote}
{\small
\begin{center}{\bf ABSTRACT}\end{center}
\vspace{0.3cm}\par\penalty10000
The electromagnetic and weak neutral current matrix elements that
enter in the analysis of parity-violating quasielastic electron
scattering are calculated using a continuum nuclear shell model.  New
approximations to the on-shell relativistic one-body currents and
relativistic kinematics for use in such models are developed and
discussed in detail. Results are presented for three closed-shell
nuclei of interest: $^{16}$O, $^{40}$Ca and $^{208}$Pb. The current
work concludes with a study of the sensitivity of the resulting
parity-violating asymmetries to properties of the nucleon form factors
including the possible strangeness content of the nucleon. }

\end{quote}

\vfill

\hbox to \hsize{\hrulefill}

$^*$ This work is supported in part by funds provided by the U.S.
Department of Energy (D.O.E.) under cooperative agreement
\#DE-FC01-94ER40818, in part by DGICYT (Spain) under Contract
Nos. PB92/0021-C02-01 and PB92-0927 and the Junta de Andaluc\'{\i}a (Spain),
in part by the European Commission under Contract ERBCHBICT920185,
and in part by NATO Collaborative Research Grant \#940183.

\end{titlepage}


\section{Introduction}


Following the discovery of parity violation (PV) in nuclear beta decay, nuclei
have played a central role in the investigation not only of the properties
of the weak interaction but also of hidden properties of nucleon and nuclear
structure. Prior to the discovery of weak neutral currents, investigations of
parity violation in electromagnetic nuclear processes were focused on the
study of asymmetries and polarizations in photonuclear absorption and
decays~\cite{Ade85}. The latter were understood as being caused by
the parity impurities in nuclear states induced by the charge-changing
parity-violating part of the
nucleon-nucleon interaction mediated by meson exchanges~\cite{Des80}.
This parity-mixing of nuclear states continues to be of interest and, for
example, is at present one of the motivations
of proposed experiments to search for nuclear anapole moments~\cite{Hax89}.

Soon after measurements of neutrino scattering at CERN~\cite{Has73} implied
the existence of weak neutral currents, the vector/axial-vector character
of these currents was confirmed by the
observation of parity violation in electron-deuteron scattering at
SLAC~\cite{Pre78}. These experiments initiated a new era for the subject
of parity violation in nuclei, and since then several experiments have been
undertaken or proposed to measure PV asymmetries in electron-nucleus
scattering (see Ref.~\cite{Mus94} for discussion of the present situation).
Two important experiments have been carried out so far on complex nuclei,
elastic scattering from $^{12}$C at Bates~\cite{PAS90} and quasielastic (QE)
scattering from $^{9}$Be at Mainz~\cite{Hei89}, yielding PV asymmetries
that are, within experimental
uncertainties, consistent with the Standard Model. Yet there is
ample room open for speculation, and one of the main interests at present for
PV electron-nucleus experiments lies in the possibility of
extracting information on elusive parts of the nucleon form factors.
In particular, as discussed at length in Ref.~\cite{Mus94}, there is at present
considerable interest in studying various aspects of the nucleon's
strangeness content, that is, its strangeness distributions
or form factors.

Exploiting the rich variety of nuclei allows one to explore various features
of the electroweak nucleon current and form factors complementary to those
accesible via lepton-proton scattering. For instance, the above-mentioned
experiment on $^{12}$C performed at Bates~\cite{PAS90} was
designed to filter out the isoscalar electroweak coupling with hadronic matter.
Or, as in the experiment on $^{9}$Be performed at Mainz~\cite{Hei89},
measuring parity
violation in QE electron scattering has the
advantage of involving large cross sections, in addition to that of
involving various combinations of isoscalar, isovector, electric,
and magnetic vector (and axial-vector) form factors, which can in
principle be selected by appropiate choices of nuclei and of kinematics.
The price one has to pay for this is that the nuclear many-body problem
enters the picture, and ambiguities in the description of nuclear structure
may render ambiguous the extraction of the various single-nucleon form factors.
Obviously, an unambiguous extraction of nucleon form factors
at least requires careful evaluation of the underlying mean-field-based
description of the nuclear structure involved.

In Ref.~\cite{Don92} a study was carried out of the sensitivity to nucleon form
factors (particularly to axial-vector and strangeness form factors) of PV
response functions and asymmetries in QE electron scattering
from $^{12}$C. The study was carried out within the context of
the relativistic Fermi gas model (RFG) and then later extended in
Refs.~\cite{Alb88,Alb90,Alb93,Bar94} to more sophisticated
models that incorporate specific classes of correlations and meson exchange
current (MEC) effects. In other work reported in Ref.~\cite{Hor93} an
approach arising from models based on relativistic mean field theory has
been pursued. As discussed in Ref.~\cite{Bar94}, for appropriately chosen
observables it appears to be possible to extract useful information
from measurements of the QE PV asymmetry in nuclei that bears on
issues of the strangeness content of the the nucleon and its axial-vector,
isovector current. Importantly, in the analyses to date the ``confusion''
from nuclear physics uncertainties appears not to obscure such extractions,
despite the fact that reservations might be raised concerning the
nuclear models used~\cite{Mar84}.

Indeed, the impulse approximation has been shown to be a good
approximation for QE kinematics and final-state interactions are not expected
to affect the PV asymmetries for high momentum transfers, so that the models
used previously are quite likely to be capable of yielding the critical
observables for extracting information on the single-nucleon form factors
from the integrated PV asymmetry. However, it is still important to explore
alternative approaches to the nuclear dynamics. Only if the appropriate
observables are stable with respect to making different choices for the
nuclear models is it then reasonable to expect that the single-nucleon
content can be extracted; if different nuclear models yield significantly
different results, then one must doubt the entire procedure at present,
since each model has its special merit and no one approach includes all
aspects of the problem (relativistic effects, many-body currents,
correlations of all types --- Hartree-Fock, RPA, {\it et cetera}).

Accordingly, in this work we extend the previous work by exploring the
various PV observables within the context of a continuum shell model (CSM)
description for comparison with the other approaches mentioned above.
In particular, in this work the CSM is applied to three closed-shell
nuclei, $^{16}$O, $^{40}$Ca and $^{208}$Pb where the approximate
expressions used for the currents are known to hold best (see Sect.~2).
This model has been thoroughly tested by Amaro
{\it et al.\/} \cite{Ama92,Ama94a,Ama94b}
for the computation of (parity-conserving) longitudinal and transverse
responses in the region of the QE peak. In Ref.~\cite{Ama92} a
comparison was presented with results obtained using the continuum random phase
approximation (RPA) model of Ref.~\cite{Coq88}, which has as a residual
interaction the effective finite-range interaction derived from the nuclear
matter  polarization potential of Ref.~\cite{PQW88}. It was found~\cite{Ama92}
that the CSM approach produces similar results at the $q$-values of concern
here (the results are practically undistinguishable at
$q=550\,{\rm MeV/c}$). In this past work the effects of
meson-exchange currents were also investigated and it was seen in
Ref.~\cite{Ama94a} that the net effect of MEC is very small except
in the very high energy tail of the QE response where contributions of
the two-particle emission channel dominate the $R^T$ response (as an example,
for $q=550\, {\rm MeV/c}$ this starts to be important only at
$\omega\sim 350$ MeV).
Thus the CSM is expected to be reliable as well
for the computation of PV response functions in the vicinity of the
QE peak.  The issue of relevance for the present work is how
different the various PV observables are in the CSM compared with those
that result when other models such as the ones mentioned above are
employed. Note that previous CSM modeling has been undertaken only
at the non-relativistic level and only for the case
of the EM responses~\cite{Ama94a,Ama94b};
here we include relativity in the calculation and consider
also the PV responses and asymmetries.

This paper is organized as follows. In Sect.~2 we very briefly review
the formalism for PV electron-nucleus scattering reactions based on
the material presented in a recent review article on the subject~\cite{Mus94}.
The discussion centers around several issues: (1) We review the essential
expressions for the general nuclear response functions and PV asymmetry;
(2) We consider the electroweak hadronic currents for on-shell nucleons and
drawing on exact answers and various approximations to these quantities
(whose details are presented in Appendix~A) we develop the single-nucleon
currents and their one-body consequences; (3) We summarize the
continuum shell model that forms the basis of the present work (again,
with some specifics relegated to Appendix~B and other connections to
the RFG model placed in Appendix~C). In Sect.~3 we present the results
obtained for the response functions and asymmetries for the cases of the
three closed-shell nuclei $^{16}$O, $^{40}$Ca and $^{208}$Pb. The
discussions in that section open with a re-examination of the
approximations made in the present work for relativistic aspects of
currents and kinematics. In particular, we apply these to the Fermi
gas model, where we know the exact answer ({\it i.e.,\/} the RFG),
as motivation for taking the same approach in the CSM. Following
this the PV asymmetry is explored with the aim of quantifying the
nuclear model uncertainties inherent in attempts to extract information
about the nucleon's strangeness and/or axial-vector content from
PV QE electron scattering.
Finally in Sect.~4 we summarize our conclusions.


\section{Summary of theory}


\subsection{Asymmetry and Nuclear Responses}

The general formalism for parity-violating electron scattering has
been presented in detail in previous work (see Refs.~\cite{Mus94,Mus92} and
references therein). Therefore, in this subsection we summarize only those
aspects of the reaction that are needed for the discussions that follow.
As in Ref.~\cite{Mus94},
we limit our attention to the plane-wave Born approximation (PWBA) with
single photon or $Z^0$ exchange. The processes we consider are represented
in Fig.~1. Here, an electron with four-momentum
$K^{\mu}=(\epsilon,{\bd k})$ and helicity $h$ is scattered
through an angle $\theta_e$ to four-momentum
$K'{}^{\mu}=(\epsilon',{\bd k'})$, exchanging a photon or a $Z^0$, where the
transferred four-momentum in the process is given by $Q^\mu =(K-K')^\mu=
(\omega, {\bd q})$. We follow the conventions of Ref.~\cite{BD64},
including the invariant phase space factors that are neglected
in some other work.

The basic quantity of interest here is the helicity asymmetry,
defined as the ratio involving the difference
and the sum of the electron scattering cross sections for positive and
negative incoming electron helicities.
The difference is parity-violating, while the sum is twice the
usual parity-conserving electromagnetic (EM) cross section. When considering
single-arm scattering of longitudinally polarized electrons from unpolarized
nucleons and nuclei, the leading-order PV contribution arises from
interferences between the two processes shown diagrammatically in Fig.~1. PV
effects can also come from $Z^0$ exchange alone, but these are much smaller
than the interference terms considered and thus are neglected here. Note
that the $\gamma$-exchange amplitude (a) is purely vector, whereas
the $Z^0$-exchange weak neutral current amplitude (b) has both vector (V)
and axial-vector (A) contributions. The helicity-difference asymmetry may
then be written
\begin{equation}
{\cal A} =
\frac{\displaystyle \frac{d\sigma^+}{d\Omega'd\epsilon'}
      -\frac{d\sigma^-}{d\Omega'd\epsilon'}}{%
      \displaystyle  \frac{d\sigma^+}{d\Omega'd\epsilon'}
      +\frac{d\sigma^-}{d\Omega'd\epsilon'}} =
	{\cal A}_0\frac{W_{\rm pv}}{W_{\rm em}},
\label{1}
\end{equation}
where the scale is set by the term
\begin{equation}
{\cal A}_0=\frac{G|Q^2|}{2\pi\alpha\sqrt{2}},
\label{2}
\end{equation}
with $G$ the Fermi coupling and $\alpha$ the fine structure constant.

The nuclear physics (hadronic) content in the problem is contained in the ratio
of the parity-violating responses ($W_{\rm pv}$) to the familiar
parity-conserving EM responses ($W_{\rm em}$). Within
the PWBA, these are given by
\beqa
W_{\rm em} &=& v_L R^L + v_T R^T
\label{3}
\\
W_{\rm pv} &=& v_L R^L_{AV}+v_T R^T_{AV}+v_{T'}R^{T'}_{VA},
\label{4}
\eeqa
where $L$ and $T$ denote longitudinal and transverse projections with respect
to ${\bd q}$, respectively. The whole dependence on the electron kinematics is
contained in the usual kinematic factors,
$v_L=Q^4/q^4$, $v_T=\tan^2(\theta_e/2)-Q^2/(2q^2)$ and
$v_{T'}=\tan(\theta_e/2)\sqrt{\tan^2(\theta_e/2)-Q^2/q^2}$.
The subscript $AV$ in the PV responses denotes interferences of axial-vector
leptonic currents with vector hadronic currents and the reverse
when written $VA$.

The five hadronic responses in Eqs.~(\ref{3},\ref{4})
can be obtained from the
components of the EM hadronic tensor $W^{\mu\nu}_{\rm em}$ in the
case of the two pure EM responses $R^L$ and $R^T$, and from
the hadronic tensor $W^{\mu\nu}_{\rm em/nc}$ which involves interferences of
EM and weak neutral currents (NC) for the three PV responses,
$R^{L}_{AV}$, $R^{T}_{AV}$ and $R^{T'}_{VA}$. By considering the laboratory
coordinate system with the $z$-axis along ${\bd q}$, the $y$-axis along
${\bd k}\times {\bd k'}$ and the $x$-axis in the electron scattering plane,
the various
hadronic response functions may be written as follows,
\beqa
R^L=W^{00}_{\rm em}&=&
	\sum_f\overline{\sum_i}\delta(E_f-E_i-\omega)
             |\langle f|\rho^{\rm em}|i\rangle|^2
\label{5}\\
R^T=W^{xx}_{\rm em}+W_{\rm em}^{yy}&=&
	\sum_f\overline{\sum_i}\delta(E_f-E_i-\omega)
             |\langle f|{\bd J}_\perp^{\rm em}|i\rangle|^2,
\label{6}
\eeqa
where the sum runs over all the final unobserved states and an average over
initial states is performed. The terms
$\rho^{\rm em}$ and ${\bd J}_\perp^{\rm em}$ represent the
Fourier transforms of the
EM nuclear charge and transverse current
operators, respectively. Current conservation has been used to eliminate
the $z$-component of the hadronic current.

For the PV responses we can write
\begin{eqnarray}
R^L_{AV}&=&-\frac{g_A}{2}W^{00}_{\rm em/nc}=
             -\frac{g_A}{2}\sum_f\overline{\sum_i}\delta(E_f-E_i-\omega)
             \langle f|\rho^{\rm em}|i\rangle^*
             \langle f|\left(\rho^{\rm nc}\right)_V|i\rangle
\label{7}\\
R^T_{AV}&=&-\frac{g_A}{2}\left(W^{xx}_{\rm em/nc}+W^{yy}_{\rm em/nc}\right)
\nonumber \\
&=&             -\frac{g_A}{2}\sum_f\overline{\sum_i}\delta(E_f-E_i-\omega)
             \langle f|{\bd J}_{\perp}^{\rm em}|i\rangle^*
             \cdot\langle f|\left({\bd J}_{\perp}^{\rm nc}
\right)_V|i\rangle
\label{8}\\
R^{T'}_{VA}&=&-\frac{ig_V}{2}\left(W^{xy}_{\rm em/nc}-W^{yx}_{\rm em/nc}\right)
\nonumber \\
& = &             -\frac{ig_V}{2}\sum_f\overline{\sum_i}\delta(E_f-E_i-\omega)
             \left[\langle f|{\bd J}^{\rm em}|i\rangle^*\times
             \langle f|\left({\bd J}^{\rm nc}\right)_A|i\rangle
             \right]\cdot\frac{\bd q}{q}.
\label{9}
\end{eqnarray}
Here the indices $V$ and $A$ refer to the vector and axial-vector contributions
to the hadronic neutral current. The terms $\left(\rho^{\rm nc}\right)_V$ and
$\left({\bd J}_\perp^{\rm nc}\right)_V$ are the Fourier transforms of the
NC vector nuclear
charge and transverse current operators, respectively, while
$\left({\bd J}^{\rm nc}\right)_A$ is the Fourier transform of the
axial-vector contribution in the NC operator.

The nature of the leptonic vertex appears here in the vector and axial-vector
electron couplings, $g_V$ and $g_A$. The Standard model for the electroweak
interaction at tree level is assumed in this work,
\beqa
g_V &=& -1+4\sin^2\theta_W \cong -0.092
\label{10}\\
g_A &=& 1,
\label{11}
\end{eqnarray}
and $\theta_W$ the Weinberg angle ($\sin^2\theta_W \cong 0.227$, as
discussed in Ref.~\cite{Mus94}).


\subsection{Electroweak Hadronic Currents}

As noted from Eqs.~(\ref{5})--(\ref{9}), in order to calculate the EM and PV
responses one needs to evaluate the pure EM and weak NC hadronic currents.
Following
the detailed study presented in Refs.~\cite{Mus94,Mus92} and eliminating
terms involving
$(c,b,t)$ quarks, whose contributions to nuclear matrix elements
of $(J_\mu^{\rm nc})_{V/A}$ are supressed, the nuclear vector and axial-vector
NC operators may be written as follows,
\beqa
\left(J^{\rm nc}_{\mu}\right)_V
		&=& \xi_V^{(T=1)}J^{\rm em}_{\mu}(T=1)+
                   \sqrt{3} \xi_V^{(T=0)}J^{\rm em}_{\mu}(T=0)
                   +\xi_V^{(0)}V_{\mu}^{(s)}
\label{12}\\
\left(J^{\rm nc}_{\mu5}\right)_A &=&
			\xi_A^{(T=1)}A_{\mu}^{(3)}+
                   \xi_A^{(T=0)}A_{\mu}^{(8)}
                   +\xi_A^{(0)}A_{\mu}^{(s)},
\label{13}
\eeqa
where $J_\mu^{\rm em}(T=1)/J_\mu^{\rm em}(T=0)$ are the isovector/isoscalar
EM currents, and $V_{\mu}^{(s)}\equiv\overline{s}\gamma_{\mu}s$ is
the strange quark contribution in the vector NC current. The operators
$A_\mu^{(a)}$ are given by $A_\mu^{(a)}\equiv \overline{q}
\lambda^a \gamma_\mu \gamma_5 q/2$, where $q$ represents the triplet
of quarks $(u,d,s)$, and the $\lambda^a$, $a=1...8$ are the Gell-Mann SU(3)
matrices. The strangeness content in the axial-vector NC current is given by
$A_\mu^{(s)}\equiv \overline{s}\gamma_\mu \gamma_5 s$. Finally, the
$\xi_V$'s and $\xi_A$'s are couplings determined by the underlying electroweak
gauge theory (see Refs.~\cite{Mus94,Mus92}). We use the minimal Standard Model
tree level couplings
\beq
\begin{array}{lr}
\xi_V^{(0)}=-1 &	\xi_A^{(0)}=1 \\
\sqrt{3}\xi_V^{(T=0)}=-4\sin^2\theta_W  &  \xi_A^{(T=0)}=0\\
\xi_V^{(T=1)}=2-4\sin^2\theta_W  & \xi_A^{(T=1)}=-2.
\end{array}
\label{14}
\eeq

Single-nucleon matrix elements of the EM and weak NC currents
shown previously are restricted by Lorentz covariance, together with parity
and time reversal invariance to the following forms,
\begin{eqnarray}
\langle N(P')|J^{\rm em}_{\mu}|N(P)\rangle &=&
\overline u(P')
\left[ F_1\gamma_{\mu}+i\frac{F_2}{2M}\sigma_{\mu\nu}Q^{\nu}\right]
u(P) \label{15}\\
\langle N(P')|\left(J^{\rm nc}_{\mu}\right)_V|N(P)\rangle &=&
\overline u(P')\left[ \tilde{F}_1\gamma_{\mu}+
                      i\frac{\tilde{F}_2}{2M}\sigma_{\mu\nu}Q^{\nu}
              \right] u(P) \label{16} \\
\langle N(P')|\left(J^{\rm nc}_{\mu}\right)_A|N(p)\rangle &=&
\overline u(P')\left[ \tilde{G}_A\gamma_{\mu}+
                      \frac{\tilde{G}_P}{M}Q_{\mu}
              \right]\gamma_5  u(P),
\label{17}
\end{eqnarray}
where $u(P)$ and $u(P')$ are the single-nucleon wave functions properly
normalized, $Q=P'-P$ is the four momentum transfer to the nucleon and $M$
is the nucleon mass. From
Eqs.~(\ref{12},\ref{13}) and (\ref{15})--(\ref{17}) one can write,
\beqa
\tilde{F}_a &=& \xi_V^{(T=1)}F_a^{(T=1)}\tau_3+
		\sqrt{3}\xi_V^{(T=0)}F_a^{(T=0)}+
		\xi_V^{(0)}F_a^{(s)},\kern 4mm a=1,2
\label{18}\\
\tilde{G}_a &=& \xi_A^{(T=1)}G_a^{(3)}\tau_3+
		\xi_A^{(T=0)}G_a^{(8)}+
		\xi_A^{(0)}G_a^{(s)},\kern 4mm a=A,P,
\label{19}
\eeqa
where $F_a^{(T=0,1)}$ denote the isoscalar and isovector EM
Dirac and Pauli form factors of the nucleon, the $G_a^{(3,8)}$ are the
triplet and octet axial-vector form factors, and $F_a^{(s)}$ and $G_a^{(s)}$
are the vector and axial-vector strange-quark form factors. In Eqs.~(\ref{18})
and (\ref{19}) the terms involving $\tau_3$ are isovector while the
rest are isoscalar. We will mainly use
throughout this paper the Sachs form factors defined as:
$G_E=F_1-\tau F_2$ and $G_M=F_1+F_2$, with $\tau\equiv |Q^2|/4M^2$.
Analogously one can also define
$\tilde{G}_E$ and $\tilde{G}_M$ from $\tilde{F}_1$ and $\tilde{F}_2$.
Also, as discussed in Appendix~A and Ref.~\cite{Mus94}, the pseudoscalar
contributions are absent in PV electron scattering.

In this work we follow the usual procedures of employing the on-shell
single-nucleon currents to construct one-body EM and NC current operators.
However, in contrast to some past work, we devise operators that retain
important aspects of relativity. In Appendix~A expressions for the exact
on-shell operators for use between two-component spin spinors are summarized
together with the approximations made in the present work when using these
operators within the CSM. Below we repeat these approximate expressions
for use in building the one-body nuclear current operators needed in
treating PV electron scattering; specifically, only the vector time
projections (EM and NC) and transverse vector (EM and NC) and axial-vector
(NC) projections are required. As also discussed in Appendix~A, since we
consider only doubly-closed-shell nuclei, it is also a reasonable
approximation to drop the spin-orbit contributions to the time
projections, as we do in the following. With dimensionless variables
$\kappa\equiv q/2M$ and $\tau$ defined above we then have
\begin{eqnarray}
\rho^{\rm em} &=& \frac{\kappa}{\sqrt{\tau}}G_E \label{20}\\
\left(\rho^{\rm nc}\right)_V &=&  \frac{\kappa}{\sqrt{\tau}}
        \tilde{G}_E \label{21}\\
{\bd J}^{\rm em} &=& \frac{\sqrt{\tau}}{\kappa}
          \left(G_E\frac{\bd p+\bd p'}{2M}+
                             iG_M\frac{\nsigma\times\bd q}{2M}\right)
\label{23}\\
\left({\bd J}^{\rm nc}\right)_V &=&
        \frac{\sqrt{\tau}}{\kappa}\left(\tilde{G}_E\frac{\bd p+\bd p'}{2M}+
                             i\tilde{G}_M\frac{\nsigma\times\bd q}{2M}\right)
\label{24}\\
\left({\bd J}^{\rm nc}\right)_A &=& \sqrt{1+\tau}\tilde{G}_A\nsigma,
\label{25}
\end{eqnarray}
where for the three-vector components of the current only the transverse
projections are to be employed --- the longitudinal projections of the
vector currents are given by the continuity equation and no longitudinal
projections of the axial-vector current are required in descriptions of
PV electron scattering.
Therefore, within the IA, the hadronic (one-body)
current operators are given by the following:
\begin{itemize}
\item Time Components
\begin{eqnarray}
\rho^{\rm em} &=& \frac{\kappa}{\sqrt{\tau}}
       \sum_{k=1}^A  {\rm e}^{i{\bf q\cdot r}_k}
           \left[ G_E^{p}
               \frac{1+\tau^k_3}{2}+G_E^{n}\frac{1-\tau^k_3}{2}\right]
                 \label{26}\\
\left(\rho^{\rm nc}\right)_V &=& \frac{\kappa}{\sqrt{\tau}}\sum_{k=1}^A
       {\rm e}^{i{\bf q\cdot r}_k}
       \left[ \tilde{G}_E^{p}
               \frac{1+\tau^k_3}{2}+
               \tilde{G}_E^{n}
               \frac{1-\tau^k_3}{2}
               \right].   \label{27}
\eeqa
\item Spatial Components (transverse projections only)
\beqa
{\bd J}^{\rm em} &=& -i\frac{\sqrt{\tau}}{\kappa}
	\sum_{k=1}^A\frac{{\rm e}^{i{\bf q\cdot r}_k}}{2M_k}
    \left[\left( G_M^{p}\frac{1+\tau_3^k}2+
         G_M^{n}\frac{1-\tau_3^k}2\right)
	{\bd q}\times\mbox{\boldmath $\sigma$}^k \right. \nonumber \\
&& \left.
	+\left(G_E^{p}\frac{1+\tau_3^k}2+
          G_E^{n}\frac{1-\tau_3^k}2\right)
	(2{\bd \nabla}_k+i{\bd q})\right]
\label{29}
\eeqa
\beqa
\left({\bd J}^{\rm nc}\right)_V &=& -i\frac{\sqrt{\tau}}{\kappa}
	\sum_{k=1}^A\frac{{\rm e}^{i{\bf q\cdot r}_k}}{2M_k}
   \left[\left(\tilde{G}_M^{p}\frac{1+\tau_3^k}2+
         \tilde{G}_M^{n}\frac{1-\tau_3^k}2\right)
	{\bd q}\times\mbox{\boldmath $\sigma$}^k \right. \nonumber \\
&&\left.
	+\left(\tilde{G}_E^{p}\frac{1+\tau_3^k}2+
          \tilde{G}_E^{n}\frac{1-\tau_3^k}2\right)
	(2{\bd \nabla}_k+i{\bd q})\right]
\label{30}
\eeqa
\beq
\left({\bd J}^{\rm nc}\right)_A = \sqrt{1+\tau}\sum_{k=1}^A
		{\rm e}^{i{\bf q\cdot r}_k}
   	\left[\tilde{G}_A^{p}\frac{1+\tau_3^k}2+
         	\tilde{G}_A^{n}\frac{1-\tau_3^k}2
   		\right] \mbox{\boldmath $\sigma$}^k,
\label{31}
\eeq

\end{itemize}
where ${\bf r}_k$ is the position of the $k$-th nucleon,
while $\nsigma^k$  and $\tau_3^k$ are the Pauli
spin and (third component) isospin matrices for the $k$-th
nucleon.
As is standard we refer to the contributions proportional to
$G_M$ and $\tilde{G}_M$ as magnetization contributions, and to those
proportional to $G_E$ and $\tilde{G}_E$ as convection contributions.


\subsection{Nucleon Form Factors}

As seen in the previous section, the EM and NC current
operators are given in terms of the familiar EM form factors,
$G_{E}^{p,n}$ and $G_{M}^{p,n}$, and the weak NC ones,
$\tilde{G}_E^{p,n}$, $\tilde{G}_M^{p,n}$ and
$\tilde{G}_A^{p,n}$. Note, however, that by using
Eqs.~(\ref{18},\ref{19}) one can express these weak NC form factors in
terms of the purely EM ones $G_{E,M}^{p,n}$, the axial-vector cases,
$G_A^{p,n}$, and the form factors $G_{E,M,A}^{(s)}$ that enter when
the nucleon has nonzero $s\overline{s}$ strangeness components. At tree
level we obtain for the nucleon weak form factors the following
expressions (see Ref.~\cite{Mus94}):
\beqa
\tilde{G}_{E,M}^{p,n} &=&
		(1-4\sin^2\theta_W)G_{E,M}^{p,n}-G_{E,M}^{n,p}
			-G_{E,M}^{(s)} \label{32}\\
\tilde{G}_A^{p,n} &=&
		\mp 2G_A^{(3)}+G_A^{(s)}.
\label{33}
\eeqa

{}From these results one easily sees that the strangeness form factors
have an isoscalar character. We must also have $G_E^{(s)}=0$ for
$Q^2=\omega^2-q^2=0$, since the nucleon has no net strangeness, whereas
the $\tau=0$ values of the remaining strangeness form factors may be
nonzero at $Q^2=0$. In the case of the axial-vector weak neutral
current form factor $\tilde{G}_A$, all of the isoscalar dependence is
contained in the strangeness form factor $G_A^{(s)}$, while the remaining
term $G_A^{(3)}$ is purely isovector ($T=1$). Note that this result is only
valid at lowest order in the Standard Model where we
have $\xi_A^{(T=0)}=0$. Therefore, in the absence of strange quarks,  the
isoscalar  axial-vector form factor $\tilde{G}_A^{(T=0)}$ would vanish at
tree-level (for a more detailed discussion of this subject
see Refs.~\cite{Mus94,Mus92}). There exist different conventions for the
NC couplings;
the inter-relationships between them are presented in Ref.~\cite{Mus94}.

In order to evaluate potential PV electron scattering experiments, it is
mandatory to
characterize the $Q^2$-dependence of the various nucleon form factors. In
Sect.~3.3 we present a detailed study of the sensitivity shown by the
asymmetries to specific aspects of the nucleon
form factors. In particular, we compare the results obtained by
calculating the EM form factors using the so-called Galster
parameterization (see Refs.~\cite{Mus94,Don92,Gal71}):
\beqa
G_E^p &=& G_V^D \label{para1}\\
G_M^p &=& \mu_p G_V^D \label{para2}\\
G_E^n &=& -\mu_n \tau G_V^D\xi_n \label{para3}\\
G_M^n &=& \mu_n G_V^D, \label{para4}
\eeqa
with $G_V^D(\tau)=[1+\lambda_V^D \tau]^{-2}$, $\xi_n(\tau)=
[1+\lambda_n \tau]^{-1}$, $\lambda_V^D=4.97$ and $\lambda_n=5.6$. In
Sect.~3.3 we consider variations of $\lambda_n$.

Following Refs.~\cite{Mus94,Mus92} the isovector axial-vector form
factor is given by
\beq
G_A^{(T=1)}=-g_A G_A^D,
\label{34}
\eeq
with $g_A=1.262$, and
$G_A^D(\tau)=[1+\lambda_A^D\tau]^{-2}$ is the axial-vector dipole form
factor. In this work we have assumed the value $\lambda_A^D=3.32$
corresponding to a dipole mass of 1 GeV. Effects due to a renormalization
of $g_A$ can be quite significant in neutral current processes. In particular,
in the case of electron scattering, uncertainties in theoretical estimates
of higher-order contributions may be on the order of $\pm 10\%$ (for a more
complete discussion of these issues see Refs.~\cite{Mus94,Mus92}). Again, in
Sect.~3.3 we return to consider variations in the magnitude of $g_A$.

Finally, generalizing the Galster parameterization,
the (purely isoscalar) strange form factors can be written as
\beqa
G_E^{(s)} &=& \rho_s\tau G_V^D \xi_E^{(s)} \label{35}\\
G_M^{(s)} &=& \mu_s G_V^D \xi_M^{(s)} \label{36}\\
G_A^{(s)} &=& g_A^{(s)} G_A^D \xi_A^{(s)}. \label{37}
\eeqa
The values chosen for $\rho_s$ and $\mu_s$ will be
discussed in Sect.~3.3, whereas the remaining form factor $G_A^{(s)}$ will be
ignored in the present work as its effect in QE electron scattering is
expected to be very small~\cite{Mus94}.
One should note that the choice given by Eqs.~(\ref{35})--(\ref{37}) is
somewhat arbitrary, except for the $\tau \rightarrow 0$ behaviour of
$G_E^{(s)}(\tau)$ dictated by the fact that the nucleon has no net
strangeness. Given that no experimental information at all exists on
$G_{E,M}^{(s)}$, while very little is known about $G_A^{(s)}$ (see
Ref.~\cite{Mus94}),
Eqs.~(\ref{35})--(\ref{37}) simply show a straightforward extension of the pure
EM form factor equations to characterize the present lack of
knowledge in these form factors. Results of future measurements may
require a different choice of parameterization. In this work, we describe
the leading, non-trivial $Q^2$-dependence of $G_{M,E,A}^{(s)}$ via the
parameters $\mu_s$, $\rho_s$ and $g_A^{(s)}$, respectively, and possible
deviations of the high-$\tau$ dipole fall-off by the functions
$\xi^{(s)}_{E,M,A}$ that show a $\tau$-dependence in the form
$\xi^{(s)}_{E,M,A}(\tau)=(1+\lambda^{(s)}_{E,M,A} \tau)^{-1}$. The
parameters $\lambda_{E,M,A}^{(s)}$ are, as yet, unconstrained. Our focus
in the present work is entirely on the region where $\tau<0.3$.

\subsection{Nuclear Shell model}

As has been already mentioned in previous sections, in this paper we restrict
ourselves to the Impulse Approximation (IA), {\it i.e.,} we only consider
one-body current operators. The study of MEC will be
presented in a forthcoming publication. In this context, apart from
different prescriptions for the nucleon
form factors, one must also model the nuclear structure in obtaining
the nuclear matrix elements that enter in the various EM and PV response
functions. In this subsection we briefly discuss the nuclear structure model
used to describe the ground and excited states of the nuclei under
consideration. In this work we restrict our attention to nuclei with
doubly-closed shells, $^{16}$O, $^{40}$Ca and $^{208}$Pb, taking the initial
states $|i\rangle \equiv |0\rangle$ to be given simply by the degenerate
ground-state wave functions
of such nuclei, whereas the final states $|f\rangle$ are represented by
1p--1h excitations (within the IA).

In the following we summarize the basic features of the CSM.
The particle and hole, proton and neutron wave functions are solutions of the
radial Schr\"odinger equation with a mean potential of Woods-Saxon
type
\beq
V(r)=\frac{-V_0}{1+\exp{(r-R_0)/a_0}}+\frac{1}{m_{\pi}^2}\frac{1}{r}
     \frac{d}{dr}\left(\frac{-V_{ls}}{1+\exp{(r-R_{ls})/a_{ls}}}\right)
     {\bd l}\cdot\nsigma+V_{\rm Coul},
\label{x1}
\eeq
where $m_{\pi}$ is the pion mass, and $V_{\rm Coul}$ is the
Coulomb potential of a uniform  charge distribution
with  radius $R_{\rm Coul}$ and charge $(Z-1)e$. Obviously, the Coulomb
potential only interacts in the case of protons.
The parameters are fixed to have a good description of
the empirical single-particle energies near the Fermi level
and of the experimental ground-state charge density. We take $R_{\rm
Coul}=R_0$.
The rest of parameters are given in Table~1 and have been taken from
Refs.~\cite{Ama94b,Ama94c,Rin78}.
In previous work~\cite{Ama92,Ama94a,Ama94b} we have analyzed the
sensitivity of the $R_L$ and $R_T$ responses to variations of these parameters.

The asymptotic form of the particle radial wave functions
of momentum $k=\sqrt{2M\epsilon}$ and angular momenta $l,j$ is
\beq
R_{\epsilon lj}(r)\sim\sqrt{\frac{2M}{\pi k}}
\frac{1}{r}\sin(kr-\eta\log2kr-l\frac{\pi}{2}+\sigma_l+\delta_{lj}),
\label{x5}
\eeq
where $\delta_{lj}$ is the nuclear phase shift, $\sigma_l$ is the
Coulomb phase shift (which is absent for neutrons),
and $\eta$  is the Coulomb parameter $M(Z-1)e^2/k$ for protons and zero
for neutrons;
the normalization of the wave functions is then
\beq
\langle R_{\epsilon lj}|R_{\epsilon' lj}\rangle=\delta(\epsilon-\epsilon').
\label{x6}
\eeq
Special care is needed when the energy is small or very large,
when the angular momentum $l$ is very large, and when the charge $Z$
is very large, since then the computation of the Coulomb wave functions
is known to be somewhat delicate.

In this approach, since we only consider inclusive electron scattering,
the nuclear final state is taken to be an eigenstate of the
total angular momentum of the many-particle system, {\it i.e.,}
\beq
|f\rangle \equiv
|\alpha,J\rangle=|p,h^{-1};J\rangle=
\left[a^{\dagger}_p\otimes b_h^{\dagger}\right]_J|0\rangle,
\eeq
where $a_p^{\dagger}$ creates a particle in the continuum, namely state
$|p\rangle=|\epsilon_p l_p j_p\rangle$ with
$\epsilon_p>0$, while $b_h^{\dagger}$ creates a hole in the
core $|0\rangle$ corresponding to destroying particle state
$|n_h l_h j_h\rangle$. Then, for a fixed value of the excitation
energy the nuclear excited state is calculated by summing over all of the
particle-hole
pairs ($\sum_\alpha$) that are allowed by the angular momentum selection rules
and energy conservation. The calculation of the responses requires the sum
over all the excited states (see Refs.~\cite{Ama92,Ama94a,Ama94b,Ama93b} for
details concerning the calculations).

When considering initial and final nuclear states with good angular momentum
quantum numbers it is convenient to expand the hadronic currents in terms
of multipole projections of the charge and transverse three-vector current
operators:
\beqa
\rho^{\rm em} &=&
	\sqrt{4\pi}\sum_{J\ge 0}(-i)^J\hat{J}
		M^J_0 \label{38}\\
J^{\rm em}_m &=&
	-\sqrt{2\pi}\sum_{J\ge 1} (-i)^J \hat{J}
		\left[T^{EJ}_m+ m T^{MJ}_m\right],
\label{39}
\eeqa
with $m=\pm 1$. We use the notation $\hat{J}\equiv \sqrt{2J+1}$.
Note that the longitudinal component of the three-current
can be obtained from the charge current by using current conservation.

The above multipole decompositions can be applied to the EM
currents, $\rho^{\rm em}$, ${\bd J}^{\rm em}$, as well as to the vector and
axial-vector weak neutral currents, $(\rho^{\rm nc})_V, ({\bd J}^{\rm nc})_V,
(\rho^{\rm nc})_A, ({\bd J}^{\rm nc})_A$. The corresponding Coulomb and
transverse electric and magnetic multipole operators are given as usual by
\beqa
M^{J}_m &=&
		\int d{\bd r} j_J(qr) Y_J^m(\hat{{\bd r}})\rho({\bd r})
\label{40}\\
T^{EJ}_m &=&
	\frac{1}{q}\int d{\bd r}\left\{ {\bd \nabla}\times
	\left[j_J(qr){\bd Y}_{JJ}^m(\hat{{\bd r}})\right]\right\}
		\cdot {\bd J}({\bd r}) \label{41}\\
T^{MJ}_m &=&
	\int d{\bd r}j_J(qr){\bd Y}_{JJ}^m(\hat{{\bd r}})
        \cdot {\bd J}({\bd r}),
\label{42}
\eeqa
with $j_J(qr)$ the spherical Bessel functions, $Y_J^m(\hat{{\bd r}})$ the
spherical harmonics and ${\bd Y}_{JJ}^m(\hat{{\bd r}})$ the vector spherical
harmonics. Here $\rho({\bd r})$ and ${\bd J}({\bd r})$ stand for any of
the nuclear EM, vector NC and axial-vector NC
current operators.

Introducing the above multipole analysis in the general expressions for the
response functions (Eqs.~(\ref{5})--(\ref{9})) and making use of the
Wigner-Eckart theorem,
it is straightforward to obtain the following expressions:
\beqa
R^L(q,\omega)&=&4\pi\sum_{J\ge 0}\sum_{\alpha}
 \delta(E_{\alpha}-E_0-\omega)
 |\langle\alpha,J\|M^J_{\rm em}\|0\rangle|^2
\label{43}\\
R^T(q,\omega)&=&4\pi\sum_{J\ge 1}\sum_{\alpha}
 \delta(E_{\alpha}-E_0-\omega)
 |\langle\alpha,J\|T^{EJ}_{\rm em}+iT^{MJ}_{\rm em}\|0\rangle|^2
\label{44}\\
R^L_{AV}(q,\omega)&=&-2\pi g_A\sum_{J\ge 0}\sum_{\alpha}
 \delta(E_{\alpha}-E_0-\omega)
 \langle\alpha,J\|M^J_{\rm em}\|0\rangle^*
 \langle\alpha,J\|\left(M^J_{\rm nc}\right)_V\|0\rangle
\label{45}\\
R^T_{AV}(q,\omega)&=&-2\pi g_A\sum_{J\ge 1}\sum_{\alpha}
\delta(E_{\alpha}-E_0-\omega)
\left[ \langle\alpha ,J\|T^{EJ}_{\rm em}\|0\rangle^*
        \langle\alpha ,J\|\left(T^{EJ}_{\rm nc}\right)_V\|0\rangle
\right. \nonumber \\
 &+ & \left. \langle\alpha,J\|T^{MJ}_{\rm em}\|0\rangle^*
         \langle\alpha,J\|\left(T^{MJ}_{\rm nc}\right)_V\|0\rangle \right]
\label{46}\\
R^{T'}_{VA}(q,\omega)&=&-2\pi g_V\sum_{J\ge 1}\sum_{\alpha}
 \delta(E_{\alpha}-E_0-\omega)
\left[ \langle\alpha,J\|T^{EJ}_{\rm em}\|0\rangle^*
        \langle\alpha,J\|\left(T^{MJ}_{\rm nc}\right)_A\|0\rangle
\right. \nonumber\\
& +& \left.  \langle\alpha,J\|T^{MJ}_{\rm em}\|0\rangle^*
         \langle\alpha,J\|\left(T^{EJ}_{\rm nc}\right)_A\|0\rangle \right],
\label{47}
\eeqa
where we have used the notation,
$M^{J}_{\rm em}$, $T^{EJ}_{\rm em}$ and $T^{MJ}_{\rm em}$
for the EM multipole operators,
$\left(M^{J}_{\rm nc}\right)_V$, $\left(T^{EJ}_{\rm nc}\right)_V$ and
$\left(T^{MJ}_{\rm nc}\right)_V$ for
the multipole operators associated with the vector NC
four-current, and $\left(T^{EJ}_{\rm nc}\right)_A$ and
$\left(T^{MJ}_{\rm nc}\right)_A$ for
the ones corresponding to the transverse axial-vector NC current.

Note that the Coulomb and electric operators have parity
$(-1)^J$ in the case of EM and weak NC vector
currents, but they have parity $(-1)^{J+1}$ for the axial-vector
current. On the other hand, the magnetic operator has parity
$(-1)^{J+1}$ for the EM and vector NC currents, and
$(-1)^J$ for the axial-vector one.

Explicit expressions for the reduced matrix elements appearing in
Eqs.~(\ref{43}--\ref{47}) are given in
Appendix~B. It is worth pointing out that the following relationships hold
for the approximate current matrix elements:
\beqa
\langle \alpha, J\|\left(T^{EJ}_{\rm nc}\right)_A\|0\rangle &=&
	\sqrt{1+\frac{1}{\tau}}\frac{\tilde{G}_A}{G_M}
	\langle \alpha, J\|T^{MJ}_{\rm em}\|0\rangle_{\rm mag} \nonumber\\
&=& \sqrt{1+\frac{1}{\tau}}\frac{\tilde{G}_A}{\tilde{G}_M}
	\langle \alpha, J\|\left(T^{MJ}_{\rm nc}\right)_V\|0\rangle_{\rm mag}
\label{48}\\
\langle \alpha, J\|\left(T^{MJ}_{\rm nc}\right)_A\|0\rangle &=&
	\sqrt{1+\frac{1}{\tau}}\frac{\tilde{G}_A}{G_M}
	\langle \alpha, J\|T^{EJ}_{\rm em}\|0\rangle_{\rm mag} \nonumber\\
&=& \sqrt{1+\frac{1}{\tau}}\frac{\tilde{G}_A}{\tilde{G}_M}
	\langle \alpha, J\|\left(T^{EJ}_{\rm nc}\right)_V\|0\rangle_{\rm mag},
\label{49}
\eeqa
where here ``mag'' denotes the magnetization current.
Theoretically the sums in Eqs.~(\ref{43})--(\ref{47}) extend to
$J_{max}=\infty$, while in practice we add together only a finite
number of multipoles. In general we use
a value of $J_{max}$ that is greater than that needed for convergence,
$J_{conv}$, specifically the values given in Table~2.

We must take into account the resonant part of the excitation spectrum
at low energy. In addition we need to integrate the responses
to compute the sum-rule asymmetry given in Eq.~(\ref{r1}). Thus we need to
compute the responses with a small step size, from emission threshold up
to a sufficiently high energy $\omega_{max}$ that the responses become
negligible compared with their peak values. In particular for $q=$ 300, 500 and
700 MeV/c we use as step sizes 1, 2 and 5 MeV and $\omega_{max}=$ 200, 350
and 500 MeV, respectively.

At low $\omega$ resonances in the Woods-Saxon potential
produce rapid oscillations and sharp peaks in the responses, whose overall
contribution we take into account by convoluting
the responses  with a Gaussian weight function
$f(\omega)\propto\exp(-\omega^2/\Gamma^2)$ of width $\Gamma$ (in all cases
we have used $\Gamma=5$ MeV).
So what we show in the plots is not actually the shell model
response $R_{\rm SM}(q,\omega)$, but the ``smoothed'' response
$R_{\Gamma}(q,\omega)$, given by
\beq
R_{\Gamma}(q,\omega)\equiv\int_{-\infty}^{\infty}
    d\omega' f(\omega-\omega')R_{\rm SM}(q,\omega').
\label{x9}
\eeq

In the next section we compare the results obtained within the present
model to those obtained with the RFG. To facilitate the
discussion there, we summarize in Appendix~C the expressions for the five
response functions within the relativistic Fermi Gas Model together with
the non-relativistic limits.


\section{Results}


In this section we present and study in detail the results
obtained for the response functions and asymmetry defined in
previous section and appendices of the present paper. An important goal
in this work has been to
explore the sensitivities that the various observables display to
different nuclear models, as well as to properties of the
single-nucleon form factors. Concerning the former,
our motivation has been to investigate whether or not
shell effects show up in the observables under consideration and
to explore these nuclear model dependences using an approach that
differs from that employed in past work in this area.
In particular, as discussed above, we have argued that the CSM augmented by
the incorporation of an approximate relativistic one-body current and
relativistic kinematics should provide a reasonable model for inclusive
medium-energy QE electron scattering. Our aim in the present
section is to present results that support this approach and then to proceed
to investigate aspects of the single-nucleon structure physics as
revealed in PV QE electron scattering.

To set a clear scale for the model dependences (and therefore the
ambiguities) in the response functions and PV asymmetry, we
have performed calculations for different values of the momentum
transfer ranging from $q=300$ MeV/c up to 1000~MeV/c using the CSM
described in Sect.~2.4. In so-doing
we deal with situations in which non-relativistic
calculations are expected to be reliable, and others in which a
relativistic analysis of the reaction mechanism is mandatory.
Results for three closed-shell nuclei, $^{16}$O, $^{40}$Ca and $^{208}$Pb,
are presented. Naturally, the PWBA used in this work is not directly
applicable in modeling the response functions for nuclei as heavy as
lead. In such high-$Z$ cases the electron distortion can be taken into
account by obtaining numerical solutions to the Dirac equation with
a Coulomb potential. In the present paper our main focus is placed on
the PV asymmetry which, being proportional to a ratio of helicity
difference and sum cross sections, is less affected by distortion.
Accordingly, for the model-to-model comparisons made here the PWBA should
prove adequate.

Throughout this work, the nucleon's EM form factors are taken to be given by
the simple parameterizations discussed in Sect.~2.3. Since it is expected that
these functions will be studied intensively at most medium- to high-energy
electron scattering facilities over the next few years, we have not
pursued any further analysis of the sensitivity of the PV asymmetry to
this aspect of the problem other than to present one figure
showing the sensitivity to variations of $G_E^n$; this issue can easily
be re-examined when new data become available from measurements of
parity-conserving polarization-dependent electron
scattering from the proton, deuteron and $^3$He.

We begin by reproducing the observables in the RFG model studied in
previous work~\cite{Don92,Alb88,Alb90,Alb93,Bar94} (see Appendix~C) to
quantify further the
approximations made in our present treatment of the CSM. Importantly,
we show below that the current developed in Appendix~A provides an
excellent approximation to the exact on-shell results when used within
the context of the Fermi gas model, confirming those past studies. The
same approach is then used in the CSM to proceed to an investigation
of the nuclear model dependences expected for PV observables.

\subsection{Quasielastic responses in the RFG model}

The RFG model has been summarized in Appendix~C and several approximations
to the single-nucleon current discussed in Appendix~A. In the present
subsection we present results using this model to help in assessing the
quality of those approximations for use later in the CSM.

In Fig.~2 we show the EM response functions $R^L$ and $R^T$ for the case
of $^{40}$Ca at $q=$ 500, 700 and 1000 MeV/c. The solid curves result from
using the fully relativistic model obtained with the exact on-shell
current operators given in Eqs.~(\ref{sn12}--\ref{sn17}), {\it i.e.,\/} using
the formalism presented in Ref.~\cite{Alb88}. Those shown as long-dashed
curves correspond to making the approximations for the
current operators given in Eqs.~(\ref{20}) and (\ref{23}). Moreover in
this latter case we have begun with the non-relativistic Fermi gas model for
the rest of the problem ({\it i.e.,\/} other than the currents) and proceeded
to make the $\lambda\equiv\omega/2M\rightarrow\lambda(1+\lambda)$ replacement
discussed at
the end of Appendix~C, accounting approximately for relativistic kinematics
in the energy-conserving $\delta$-function. As is clearly seen by comparing the
solid and long-dashed curves, and in accord with past
work \cite{Alb90,Alb93,Bar94}, when both approximations are made the
agreement is excellent, verifying that the use of the particular
approximate currents and kinematics assumed here yields results that
differ by only a few percent from the exact RFG results.

Also shown in Fig.~2 (as short-dashed curves) are responses having the
relativized kinematics, but now using non-relativistic currents. In
particular for these the factors $\kappa/\sqrt{\tau}$ and $\sqrt{\tau}/\kappa$
in Eqs.~(\ref{20}) and (\ref{23}), respectively, are set to unity. Without the
relativistic effects in the currents the results for $R^L$ ($R^T$) are too
small (large), as expected. We note that conventional treatments of the
single-nucleon current (see, for example, Eqs.~(D4) and (D5) of
Ref.~\cite{Def66}) often take $\omega\ll q$, in which case $\kappa\cong
\sqrt{\tau}$. Moreover, as discussed in Appendix~A, the non-relativistic
expansions are made not only with respect to $\eta$, but also $\kappa$,
$\lambda$ and $\tau$. Figure~2 demonstrates the magnitude of the error
to be expected for QE scattering if one makes such approximations.

Finally in Fig.~2 are also shown results where the replacement
$\lambda\rightarrow\lambda(1+\lambda)$ is not made (dot-dash curves with the
relativistic corrections to the current, dotted curves without). With
non-relativistic kinematics the QE peak occurs near $\omega=q^2/2M$
($\lambda=\kappa^2=\tau(1+\tau)$), whereas with relativistic kinematics
it occurs near $\omega=|Q^2|/2M$ ($\lambda=\tau$), namely, at a smaller
value. As discussed in Ref.~\cite{Alb88}, the relativistic model yields
a smaller width to the QE response than does the non-relativistic model.
Also, the four-momentum transfer is decreasing as $\omega$
increases for fixed $q$ and consequently the single-nucleon form factors
(those that are simply proportional to dipole form factors) increase with
increasing $\omega$. This is the underlying reason for the large
enhancements seen in the non-relativistic results shown in the figure.

The importance of maintaining relativistic kinematics in the energy-conserving
$\delta$-function is obvious from these results. Indeed it would be a very
poor approximation to ignore either of the ``relativizing'' steps made in
the present work. Clearly as the momentum transfer reaches the
medium-energy domain of interest in this work the effects of relativity
in currents and kinematics (at least) are rather important. Typically
one would incur significant error in using the completely non-relativistic
FG model and thus, by extension, one must doubt any other model which
does not incorporate the effects discussed in the present work.
The CSM model discussed below has been developed with this in mind: we
use only the relativized current operators and always make the
replacement $\omega\rightarrow\omega(1+\omega/2M)$ in the
energy-conserving $\delta$-functions. We now turn to the results
obtained using that model.

\subsection{Quasielastic responses in the CSM}

An important aspect in the study of nuclear reactions is to
determine as precisely as we can the nuclear physics dependences
inherent in the observables. On the one hand, as in much of
the present work, we may be interested in minimizing such dependences
with the hope of extracting information about the nucleon's
various form factors; on the other hand, it may be that certain
observables contain unusually large sensitivity to some specific
issue of interest in studying the nuclear many-body problem.
The focus in the present work is largely to address the former
of these two points of view specifically for the PV electron scattering
observables. In this regard, the study of ratios of responses such as
the PV asymmetry is favored, since they usually turn out to be less sensitive
to the nature of the underlying dynamical assumptions made when compared with
the response functions (or cross sections) themselves.

In particular, here we compare the results of calculations using
the CSM with well parameters given in Table~1
with those of the RFG model for the response functions, the
PV asymmetry and the sum-rule asymmetry, the last being defined below
following previous studies in this area~\cite{Bar94}. For the RFG
results shown here we use as Fermi momenta the following: $p_F=$ 215 or
225 MeV/c for $^{16}$O, 235 or 245 MeV/c for $^{40}$Ca and 230 or
250 MeV/c for protons in $^{208}$Pb.
In each case these have been chosen as representative for the three
nuclei, with a smaller value that is favored somewhat when ground-state
properties are emphasized
and a larger value that yields rough agreement when the
RFG model predictions are compared with experiment (this presumably
arises since QE scattering at intermediate energies involves
the ground state and a rather excited particle-hole state, both of
which the model must recognize). For $^{208}$Pb we show
results using two different assumptions for the relationship between the
proton and neutron Fermi momenta, $(1)$ with $p_{Fn} = (N/Z)^{1/3} p_{Fp}=$
265 or 288 MeV/c for the two cases,
corresponding to keeping the volume of the neutron and proton gases equal,
and $(2)$ with $p_{Fn}=p_{Fp}=$ 250 MeV/c, corresponding to equal densities
for the two gases. As usual we take $p_{Fn}=p_{Fp}$ for $^{16}$O and $^{40}$Ca.
In the next subsection we show results for the PV asymmetry that indicate the
very weak dependence of this observable on the Fermi momenta chosen,
confirming past work \cite{Don92}.

The resulting EM and PV response functions are displayed in Figs.~3--8
for $q=$ 300, 500 and 700 MeV/c (the PV results given here are shown
in the absence of strangeness; see the later discussion in the next
subsection). Specifically, in Fig.~3 we show the EM response functions for
$^{16}$O, followed in Figs.~4--8 with results for the complete set of
five electroweak inclusive responses for the cases of $^{40}$Ca and
$^{208}$Pb. The inter-comparisons of $R^L$ and $R^T$ for oxygen and
calcium are typical in that the full sets of responses in these two
cases are very similar. Thus, for brevity we have focused on calcium and
lead to show the nuclear model dependences obtained in going from light
to heavy nuclei while omitting the PV responses for oxygen.

{}From the results in Figs.~3--8 we observe the following:

\begin{enumerate}
\item Overall, the RFG yields a reasonable
description of the responses when the momentum transfer is high enough.
The region where the RFG is Pauli-blocked (the $q=$ 300 MeV/c case in
the figures presented here) is where the agreement is the worst, as expected,
since the RFG is known to become a poor approximation in this
regime.

\item On closer examination, one sees that the CSM produces tails extending
below and above the response region where the RFG is nonzero. Again, this is
expected, since the latter model has a sharp cutoff to its momentum
distribution while the former has all momentum components present, these
being governed by the initial- and final-state wave functions used in the
CSM. One also sees, in going from oxygen to calcium to lead
at intermediate-to-high values of $q$, that the agreement between the two
models becomes better and better. The slight shift in peak position seen
in Fig.~3 for oxygen is seen to go away for the heavier nuclei and the
importance of the tails of the response functions produced by the CSM
decreases with increasing $A$. With regard to the latter, in effect, the
heavier the nucleus the more it appears to be dominated by volume effects,
rather than surface effects, for the inclusive responses.
At $q=300$ or 500 MeV/c and low energy we see somewhat irregular behaviour
in the shell model responses which is a consequence of the contribution
of the potential resonances in that region.
This behaviour is more pronounced in the case of $^{40}$Ca than for $^{16}$O,
and, in the case of $^{208}$Pb, it is responsible for the peak
that clearly can be seen at $q=300$ MeV/c and $\omega\cong 40$ MeV
in both $R^L$ and $R^T$.

\item The rather good agreement between the RFG and CSM responses at high $q$
is only obtained when the approximations discussed above are made. If the
currents and kinematics are not ``relativized'' as we have in this work,
but instead are taken to be their usual non-relativistic forms, then
significant differences between the RFG and (conventional) CSM will be seen
to occur.

\item When comparing the five different electroweak responses, the worst
agreement between the models is seen for $R^L_{AV}$, although even there
it is reasonably good for high $q$. It should be noted that this response
is strongly suppressed in such mean-field calculations
for the reasons discussed in Refs.~\cite{Mus94,Don92}; indeed, at
$q=$ 500 MeV/c for example, it is only a few percent of the leading PV
response $R^T_{AV}$ (and the other, $R^{T'}_{VA}$, is roughly 20\% of this).
As a consequence any differences seen here are unimportant for the PV
asymmetry. On the other hand,
especially at relatively low $q$ this is known
to be only part of the full story. In particular, it has been
argued~\cite{Bar94} that correlation effects that fall outside the
scope of the strict RFG and CSM approaches can strongly influence this
particular response at low-to-moderate momentum transfers, at the same
time only mildly effecting the other four response functions. In the next
subsection we return to this issue when discussing the role played by the
electric strangeness content of the nucleon.

\item In the figures we have shown the RFG results otained with the different
choices for the Fermi momenta discussed above. The response functions are
seen to differ somewhat, with a slight tendency to favor smaller values
for $p_F$. As we shall see immediately below, even this mild dependence
on the Fermi momenta is considerably lessened for the PV asymmetry.

\end{enumerate}

Next we turn to the PV asymmetry. As seen in Fig.~9, such ratios
are very insensitive to the choice of nuclear model made for the
dynamics. Only in kinematic regions that correspond to the tails of
the response functions do we see significant differences --- and there
the figure-of-merit for measuring the PV asymmetry is very small (see
Ref.~\cite{Mus94} for a discussion of the figure-of-merit). As a
consequence, we show only the regions where the figure-of-merit is
appreciable in the figure.

In the next subsection we shall discuss the sensitivity of PV observables
to specific aspects of the nucleon's form factors. Following Ref.~\cite{Bar94}
we shall do this using the sum-rule asymmetry (SRA) defined as:
\beq
{\cal R}(q,\theta_e)\equiv \frac{\int_0^\infty
   d\omega W_{\rm pv}(q,\omega,\theta_e)/\tilde{X}'_T} {\int_0^\infty
   d\omega W_{\rm em}(q,\omega,\theta_e)/X'_T }, \label{r1}
\eeq
where the functions $X'_T$ and $\tilde{X}'_T$ are given in
Ref.~\cite{Bar94}. These functions depend on the Fermi momentum and for
simplicity we have used only the values (in MeV/c) $p_F=$ 225 ($^{16}$O), 235
($^{40}$Ca) and 250/288 (p/n $^{208}$Pb). The dependence on $p_F$ here is
negligible and so choosing any reasonable value is an excellent approximation.
The motivation for using the specific derived quantity ${\cal R}$
is that it has been designed to be effective in removing most
of the rapid variation of the single-nucleon form factors, to the
extent that this is possible. While constructed with the RFG in mind,
however, it should be noted that it merely leads to a reduced quantity
in much the same way that dividing the nuclear cross section by the Mott
cross section leads to a form factor --- there is no loss of information,
only a re-expression of the same information. Experimentally
this quantity can be constructed from the measured
helicity-difference and helicity-sum cross sections (PV and EM,
respectively) upon dividing by the $X$'s and integrating over a range
in $\omega$. For the RFG the range of integration is naturally taken
to include the entire response, since the sharp cutoff guarantees
specific limits. In the case of the CSM, we integrate from $\omega$ at
threshold up to sufficiently large values of $\omega$ that the integrals
in the numerator and denominator of Eq.~(\ref{r1}) saturate.

In Fig.~10 we show the SRA
for $^{40}$Ca and $^{208}$Pb (the $^{16}$O results are very similar to
the former and so are not shown). The sets of curves shown are described
in the figure captions (see Fig.~4 for the basic key). Rather clearly
the nuclear model dependence is very weak, especially for
intermediate-to-high values of $q$, in that the various curves cannot
be distinguished in the figure. This weak nuclear model dependence is
essential if one wishes to use nuclei to explore properties of the
nucleon itself.

In Fig.~11 we show the CSM predictions for the SRA now for all three
nuclei, observing the near universality of this observable for a large
range of nuclei with only slight differences appearing when lead is
compared with oxygen and calcium (the last two are essentially
indistinguishable).

\subsection{Sensitivity to variations in the form factors of the nucleon}

We now turn to an assessment of the nuclear model dependences seen
using the CSM and RFG when attempting to extract information about
the isovector axial-vector and electric strangeness form factors of
the nucleon. We do this using the sum-rule asymmetry defined above.

We begin by presenting results in Fig.~12 for the SRA for $^{40}$Ca and
$^{208}$Pb with no magnetic strangeness (as in the previous figures) and
with $\mu_s$ in Eq.~(\ref{36}) set equal to $-0.35$. The latter is chosen in
accordance with the discussions of modeling the nucleon's strangeness (see
\cite{Mus94} and references therein). For comparison, as also discussed in
that reference, we note that
the SAMPLE experiment being performed at Bates to measure
backward-angle PV ${\vec e}p$ scattering is expected to have an uncertainty
of approximately $|\Delta\mu_s|\sim 0.2$. In the figure we see that
the nuclear model dependence is weaker than the dependence on
$G_M^{(s)}$ for intermediate-to-high values of $q$ in the case of an
$N=Z$ nucleus such as $^{40}$Ca ($^{16}$O is almost indistinguishable
from $^{40}$Ca). However, at low $q$ or for nuclei with $N$ significantly
larger than $Z$, such as $^{208}$Pb, the $G_M^{(s)}$ dependence is
much weaker. As discussed in Ref.~\cite{Don92}, this arises from the fact
that in the PV asymmetry this form factor is multiplied by the combination
$(ZG_M^p + NG_M^n)/A$, which is proportional to $(Z\mu_p + N\mu_n)/A= 0.44$
$(-0.06)$ for calcium (lead).

For backward-angle PV QE electron scattering the asymmetry depends
importantly on the magnetic strangeness
form factor $G_M^{(s)}$ discussed above and the isovector, axial-vector
form factor $G_A^{(T=1)}$ parametrized in Eq.~(\ref{34}). The electric
strangeness form factor $G_E^{(s)}$ is unimportant for $\theta_e$ large.
In Fig.~13 we show the dependence on $G_A^{(T=1)}$. The three sets of curves
shown correspond to results with the isovector, axial-vector coupling at
$Q^2=0$ taken to be its canonical value, $g_A=1.262$, together with results
when $g_A=1.262\pm$10\% (indicated with a $(\pm)$ in the figure). In
all cases the strangeness content of the nucleon has been neglected.
Clearly, as in Ref.~\cite{Bar94} where other nuclear models were
inter-compared in the same way, for intermediate values of momentum transfer
the nuclear model dependences are quite weak compared with the level of
variation in ${\cal R}$ caused by 10\% changes in the isovector axial-vector
coupling. Hence the conclusion found in previous work still stands that
measurements of this
observable could constrain the NC axial-vector form factor of the nucleon
with interesting precision. It is also clear from Figs.~12 and 13 that
measurements on two nuclei are needed to determine both $\mu_s$ and
$g_A$ --- PV elastic scattering from the proton and PV QE scattering
from nuclei such as those discussed in the present work or the
deuteron discussed in Ref.~\cite{Had92} would provide the required
information.

Turning finally to a discussion of forward-angle PV QE electron
scattering we focus on the dependences in the SRA on $G_E^n$, $G_M^{(s)}$
and $G_E^{(s)}$ ($G_A^{(T=1)}$ is unimportant for $\theta_e$ small).
In Fig.~14 ${\cal R}$ is shown for three values of the Galster
parameter $\lambda_n$ in Eq.~(\ref{para3}), corresponding to $\pm$10\%
variation in $G_E^n$ at its peak value. This sets a scale for the
uncertainty expected in the SRA at forward angles, being typical of
the hoped-for precision in determining $G_E^n$ from polarization
measurements of parity-conserving $e^2$H and $e^3$He scattering. The
$G_M^{(s)}$ sensitivity has been shown in Fig.~12 and the dependence
on $G_E^{(s)}$ is shown in Fig.~15. In the latter three models are employed:
(I) no electric strangeness; (II) $G_E^{(s)}$ with $\rho_s=-3$,
$\lambda_E^{(s)}=5.6$ in Eq.~(\ref{35});
and (III) $G_E^{(s)}$ with $\rho_s=-3$, $\lambda_E^{(s)}=0$. Model (II)
corresponds to ``modest electric strangeness'', while model (III) has
``large electric strangeness'' (see Ref.~\cite{Mus94}). Clearly when
the interest is to see effects from electric strangeness the momentum
transfer cannot be too low, since the single-nucleon form factor
$G_E^{(s)}$ vanishes as $|Q^2|\rightarrow 0$ (see Eq.~(\ref{35})). In the
figure we have chosen to present results for $^{40}$Ca and $^{208}$Pb at
$q=$ 300, 500 and 700 MeV/c; again the $^{16}$O and $^{40}$Ca results are
rather similar and so the former are not displayed. We see that at
forward angles and high momentum transfer
the sum-rule weighted asymmetry shows clear signatures of the electric
strangeness, while at lower values of $q$ the nuclear model dependences
become important enough  for the case of $^{208}$Pb to obscure the
$G_E^{(s)}$ effects. Comparing Figs.~12, 14 and 15 it is clear that,
given PV ${\vec e}p$ and QE information at backward angles to determine
$G_M^{(s)}$ and $G_A^{(T=1)}$ as well as possible, forward-angle PV QE
scattering at high $q$ should be capable of determining $G_E^{(s)}$.
For low $q$ the situation is much less clear. For $^{40}$Ca the
results in Fig.~15 would appear to indicate at least that models II and III
could be distinguished from model I for the electric strangeness.
However, it is already clear from the results presented in Ref.~\cite{Bar94}
that correlations beyond the mean field play a dominant role at low $q$.
In recent work~\cite{Bar95} these effects have been explored in depth
and, while the determining factor in the SRA at forward angles and
low-$q$, they become sufficiently weak at intermediate-to-high values of
$q$ that the above conclusions concerning the dominant sensitivity to
$G_E^{(s)}$ remain valid.


\section{Conclusions}


In conclusion, in this work we have (1) developed the single-nucleon vector
and axial-vector current operators for use in parity-conserving and
-violating electron scattering studies, (2) explored approximations to
these currents and to specific aspects of relativistic kinematics within
the context of the Fermi gas model, and (3) applied the same ideas to
studies of electroweak QE responses and asymmetries in the
continuum shell model.

In developing the current operators we have seen that the traditional
expansions in all dimensionless momenta provide poor approximations
to the correct on-shell answers at high momentum transfers. By making
more limited expansions in $\eta=p/M$, but not in $\kappa=q/2M$,
$\lambda=\omega/2M$ or $\tau=\kappa^2-\lambda^2$ we have obtained
approximate currents that should be appropriate in the region where
the QE responses are large. Naturally, if $\eta$ is of order
unity or large, then this scheme will be suspect; however, in such
circumstances the idea of using any on-shell current must also be questioned
and we must resort to off-shell prescriptions for the currents.

To evaluate the quality of the approximations made to the currents we have
considered the relativistic Fermi gas model where exact (model) results exist.
We have verified that the approximate currents are very successful for the
closed shell (spin-saturated) nuclei studied in this work. Furthermore,
we have verified that the non-relativistic Fermi gas model extended by
incorporating the approximate currents and as well by ``relativizing'' the
kinematics provides excellent agreement with the RFG results for the
electroweak responses. This has led us to propose that the same
approximate currents and assumption about the kinematics be invoked for
any non-relativistic model of the nuclear dynamics involved in
QE electron scattering.

In particular, we have made these assumptions in computing the QE
responses with the continuum shell model. The results are very encouraging:
the resulting electroweak response functions are rather similar to
those found with the RFG model when the momentum transfer is high
enough for the latter to be valid. Said the opposite way, if the two
assumptions concerning the currents and kinematics were not to be made
then the CSM and RFG results would be dramatically different at high $q$.

Finally, we have employed the RFG (with various choices of Fermi
momenta) and the CSM to study the PV asymmetry. We find very little
nuclear model dependence in the results, especially those relating to
the sum-rule asymmetry. This confirms the belief from other work that at high
$q$ the SRA can be used to study specific properties of the nucleon itself
without incurring too much uncertainty from the nuclear modeling.
Backward-angle PV QE scattering together with PV elastic ${\vec e}p$
scattering will serve to determine the magnetic strangeness and
isovector, axial-vector form factors of the nucleon, while forward-angle
PV QE scattering at high-$q$ may add information about the electric
strangeness form factor of the nucleon.


\subsection*{Appendix A: On-Shell Single-Nucleon Electroweak Current Operators}

The on-shell single-nucleon four-vector EM current may be written in the form:
\begin{equation}
J^\mu (P\Lambda ;P^{\prime }\Lambda ^{\prime })=\bar u(P^{\prime }\Lambda
^{\prime })\left[ F_1\gamma ^\mu +\frac i{2M}F_2\sigma ^{\mu \nu }Q_\nu
\right] u(P\Lambda ),  \label{sn1}
\end{equation}
where the incident nucleon has four-momentum $P^\mu =(E,\bd{p)}$%
, the outgoing nucleon has four-momentum $P^{\prime \mu }=(E^{\prime },\bd{p%
}^{\prime })$ and $Q^\mu =P^{\prime \mu }-P^\mu $. The spin projections are
labeled $\Lambda $ and $\Lambda ^{\prime }$ for incoming and outgoing
nucleons, respectively. As in previous work~\cite{Alb88}, the dimensionless
variables
\begin{eqnarray}
\lambda &\equiv &\omega /2M  \nonumber \\
\bd{\kappa } &\equiv &\bd{q}/2M  \nonumber \\
\tau &\equiv &\kappa ^2-\lambda ^2  \label{sn2} \\
\bd{\eta } &\equiv &\bd{p}/M  \nonumber \\
\varepsilon &\equiv &E/M=\sqrt{1+\eta ^2}  \nonumber
\end{eqnarray}
prove to be convenient.
Let us also introduce the angle $\theta $ between $\bd{\kappa }$ and $%
\bd{\eta }$. We then have the following relationships amongst the
kinematic variables:
\begin{eqnarray}
\kappa \eta \cos \theta  &=&\lambda \varepsilon -\tau \nonumber \\
\tau \left( \varepsilon +\lambda \right) ^2 &=& \kappa^2 \left( 1+\tau
+\delta^2 \right), \label{sn7}
\end{eqnarray}
where we define $\delta\equiv \eta\sin\theta$.
For reasons that will become apparent later we also define the following
quantities:
\begin{eqnarray}
\mu _1 &\equiv &\frac{\kappa \sqrt{1+\tau }}{\sqrt{\tau }\left( \varepsilon
+\lambda \right) }=\frac 1{\sqrt{1+\frac 1{1+\tau } \delta^2
}}  \label{sn8} \\
\mu _2 &\equiv &\frac{2\kappa \sqrt{1+\tau }}{\sqrt{\tau }\left( 1+\tau
+\varepsilon +\lambda \right) }=\frac{2\mu _1}{1+\frac{\sqrt{\tau (1+\tau )}}%
\kappa \mu _1}.  \label{sn9}
\end{eqnarray}

To evaluate the current in Eq.~(\ref{sn1}) we insert the appropriate $\gamma
$-matrices between the spinors
\begin{eqnarray}
\bar u(P^{\prime }\Lambda ^{\prime }) &=&\frac 1{\sqrt{2}}\sqrt{%
1+\varepsilon ^{\prime }}\chi _{\Lambda ^{\prime }}^{\dagger }\left(
1,\;-\frac 1{1+\varepsilon ^{\prime }}\bd{\eta }^{\prime }\cdot \bd{%
\sigma }\right)   \nonumber \\
u(P\Lambda ) &=&\frac 1{\sqrt{2}}\sqrt{1+\varepsilon }\left(
\begin{array}{c}
1 \\
\\
\frac 1{1+\varepsilon }\bd{\eta }\cdot \bd{\sigma }
\end{array}
\right) \chi _\Lambda ^{},  \label{sn10}
\end{eqnarray}
where $\chi _\Lambda ^{}$ and $\chi _{\Lambda ^{\prime }}^{}$ are the usual
2-component spin-$\frac 12$ spinors, with $\bd{\eta}'$ and $\varepsilon'$ the
final outgoing nucleon analogs of the unprimed quantities.
 We wish to have expressions for the single-nucleon
EM current operators $\bar J^\mu (P;P^{\prime })$ that occur inside these
latter quantities, {\em viz.}
\begin{equation}
J^\mu (P\Lambda ;P^{\prime }\Lambda ^{\prime })\equiv \chi _{\Lambda
^{\prime }}^{\dagger }\bar J^\mu (P;P^{\prime })\chi _\Lambda ^{}.
\label{sn11}
\end{equation}
We use the bar over the current in the present appendix to distinguish
an operator from its spin matrix elements; in the body of the paper
for simplicity we suppress this notation.
Writing these in the following way with an overall factor $f_0$ removed
(note that $V^{\mu}$ is not a four-vector),
\begin{eqnarray}
\bar J^\mu  &\equiv &f_0V^\mu   \label{sn12} \\
f_0 &\equiv &\frac 1{\mu _1\sqrt{1+\frac \tau {4\left( 1+\tau \right)
}{\mu _2}^2 \delta^2}},  \label{sn13}
\end{eqnarray}
the EM current operator may then be expressed in terms of the quantities
defined above,
\begin{eqnarray}
V^0 &=&\xi _0+i\xi _0^{\prime }\left( \bd{\kappa }\times \bd{\eta }%
\right) \cdot \bd{\sigma }  \label{sn14} \\
V^3 &=&\left( \lambda /\kappa \right) V^0  \label{sn15} \\
\bd{v}^{\bot } &=&\xi _1\left[ \bd{\eta }-\left( \frac{\bd{%
\kappa }\cdot \bd{\eta }}{\kappa ^2}\right) \bd{\kappa }\right] -i%
\bigg\{\xi _1^{\prime }\left( \bd{\kappa }\times \bd{\sigma }\right)
\label{sn16} \\
&&\left. +\xi _2^{\prime }\left( \bd{\kappa }\cdot \bd{\sigma }%
\right) \left( \bd{\kappa }\times \bd{\eta }\right) +\xi _3^{\prime
}\left[ \left( \bd{\kappa }\times \bd{\eta }\right) \cdot \bd{%
\sigma }\right] \left[ \bd{\eta }-\left( \frac{\bd{\kappa }\cdot
\bd{\eta }}{\kappa ^2}\right) \bd{\kappa }\right] \right\} ,
\nonumber
\end{eqnarray}
where the $\xi $'s (no spin dependence) and $\xi ^{\prime }$'s (spin
dependence) are the following:
\begin{eqnarray}
\xi _0 &=&\frac \kappa {\sqrt{\tau }}\left[ G_E+\frac{\mu _1\mu _2}{2(1+\tau
)}\delta^2\tau G_M\right]   \nonumber \\
\xi _0^{\prime } &=&\frac 1{\sqrt{1+\tau }}\left[ \mu _1G_M-\frac 12\mu
_2G_E\right]   \nonumber \\
\xi _1 &=&\frac 1{\sqrt{1+\tau }}\left[ \mu _1G_E+\frac 12\mu _2\tau
G_M\right]   \nonumber \\
\xi _1^{\prime } &=&\frac{\sqrt{\tau }}\kappa \left( 1-\frac{\mu _1\mu _2}{%
2(1+\tau )}\delta^2\right) G_M  \label{sn17} \\
\xi _2^{\prime } &=&\frac{\sqrt{\tau }}{2\kappa ^2\sqrt{1+\tau }}\sqrt{\tau
\mu _2^2+4\mu _1(\mu _2-\mu _1)-\left( \mu _1\mu _2\delta \right)
^2}G_M  \nonumber \\
&=&\frac{\lambda \sqrt{\tau }}{2\kappa ^3}\mu _1\mu _2G_M  \nonumber \\
\xi _3^{\prime } &=&\frac{\sqrt{\tau }}{2\kappa (1+\tau )}\mu_1 \mu_2 \left[
G_E-G_M\right] .  \nonumber
\end{eqnarray}
These expressions constitute exact expressions for the on-shell EM current
operator. Equation (\ref{sn15}) reflects the fact that the current is
conserved.

The vector NC analogs of the results here are simply obtained by replacing
$G_E$ by $\tilde{G}_E$ and $G_M$ by $\tilde{G}_M$.

The NC axial-vector current may be written as in Eq.~(\ref{sn1}):
\begin{equation}
J_5^\mu (P\Lambda ;P^{\prime }\Lambda ^{\prime })=\bar u(P^{\prime }\Lambda
^{\prime })\left[ {\tilde G}_A\gamma ^\mu +\frac i{2M}{\tilde G}_PQ^\mu
\right] \gamma
_5u(P\Lambda ),  \label{sn33}
\end{equation}
where the ``5'' is used to indicate axial-vector quantities (from the extra $%
\gamma _5$ above, compared with the EM current in Eq.~(\ref{sn1})) and where
${\tilde G}_A$ and ${\tilde G}_P$ are the axial and induced pseudoscalar form
factors of the nucleon, respectively. The analog of Eq.~(\ref{sn11}) becomes
\begin{equation}
J_5^\mu (P\Lambda ;P^{\prime }\Lambda ^{\prime })\equiv \chi _{\Lambda
^{\prime }}^{\dagger }\bar J_5^\mu (P;P^{\prime })\chi _\Lambda ^{},
\label{sn34}
\end{equation}
with
\begin{equation}
\bar J_5^\mu \equiv f_0A^\mu ,  \label{sn35}
\end{equation}
where $f_0$ is given in Eq.~(\ref{sn13}) (as above, here $A^{\mu}$ is not
a four-vector). Clearly the pseudoscalar
contributions are absent for transverse
(1 and 2 components) projections of the current, since
the coordinate system used has been chosen to have the momentum transfer
along the 3-direction. One finds from carrying out the procedures discussed
above for the EM (vector) current, inserting the spinors and $\gamma $%
-matrices, that the transverse projections are given by
\begin{equation}
\bd{a}^{\bot }=\left\{ \zeta _1^{\prime }\bd{\sigma }^{\bot }+\zeta
_2^{\prime }\left[ \left( \bd{\kappa }+\bd{\eta }\right) \cdot
\bd{\sigma }\right] \left[ \bd{\eta }-\left( \frac{\bd{\kappa }%
\cdot \bd{\eta }}{\kappa ^2}\right) \bd{\kappa }\right] -i\zeta
_1\left( \bd{\kappa }\times \bd{\eta }\right) \right\} {\tilde G}_A
\label{sn36}
\end{equation}
with
\begin{eqnarray}
\zeta _1 &=&\zeta _2^{\prime }=\frac 12\left( \frac{\sqrt{\tau }}\kappa
\right) \mu _1\mu _2  \nonumber \\
\zeta _1^{\prime } &=&\sqrt{1+\tau }\mu _1,  \label{sn37}
\end{eqnarray}
in parallel with Eqs.~(\ref{sn17}).

The exact results given above in Eqs.~(\ref{sn11})--(\ref{sn17})
may be used in treating EM interactions with
an on-shell nucleon. However, for applications to nuclear physics as in
the present work where
nucleons are off-shell in general it is necessary to make some
approximations. What is often done is to continue to use the above
expressions as operators inserted between nuclear wave functions, where the
latter are not on-shell plane waves but are the single-particle wave
functions for nucleons in nuclei ({\em i.e.,} in the presence of the
nuclear mean field). The coordinate space operator corresponding to $\bd{%
\eta }$ is then $-\frac i{M}\bd{\nabla }$. For typical situations the
nucleon to which the virtual photon attaches is relatively low in energy ---
the dimensionless momentum $\eta \sim \eta _F=p_F/M\sim 0.25$ yielding
dimensionless energy $\varepsilon \cong 1+\frac 12\eta ^2\sim 1+\frac
12\eta _F^2\sim 1.03.$ Accordingly the usual practice is to expand the
expressions for the EM current given above (at least) in powers of $\eta $,
retaining only the leading-order terms. Equations (\ref{sn17},\ref{sn37})
have been
cast in forms that make this easy. Let us work to linear order in $\eta $.
First, from Eqs.~(\ref{sn8}) and (\ref{sn13}) we have
$\mu _1=1+{{\cal O}}(\eta ^2)$
and $f_0=1+{{\cal O}}(\eta ^2)$, whereas it may be shown that $%
\mu _2=1+\frac 12\sqrt{\frac \tau {1+\tau }}\eta \cos \theta +{{\cal O}}%
(\eta ^2)$ by using Eqs.~(\ref{sn7}--\ref{sn9}). We then have
for the EM current operators carried to order $\eta $:
\begin{eqnarray}
\bar J^0 &=&\frac \kappa {\sqrt{\tau }}G_E+\frac i{\sqrt{1+\tau }}\left[
G_M-\frac 12G_E\right] \left( \bd{\kappa }\times \bd{\eta }\right)
\cdot \bd{\sigma }+{{\cal O}}(\eta ^2)  \label{sn29} \\
\bar J^3 &=&\left( \lambda /\kappa \right) \bar J^0  \label{sn30} \\
{\bar{\bd J}}^{\bot } &=&-\frac{\sqrt{\tau }}\kappa \left\{ iG_M\left(
\left[ \bd{\kappa }\times \bd{\sigma }\right] +\frac 1{2(1+\tau
)}\left( \bd{\kappa }\cdot \bd{\sigma }\right) \left( \bd{\kappa
}\times \bd{\eta }\right) \right) \right.  \nonumber \\
&&-\left. \left( G_E+\frac 12\tau G_M\right) \left[ \bd{\eta }-\left(
\frac{\bd{\kappa }\cdot \bd{\eta }}{\kappa ^2}\right) \bd{\kappa
}\right] \right\} +{{\cal O}}(\eta ^2),  \label{sn31}
\end{eqnarray}
employing Eqs.~(\ref{sn14}--\ref{sn16}) and noting from Eq.~(\ref{sn7}) that,
after some work,
$\kappa =\sqrt{\tau (1+\tau )}+\tau \eta \cos \theta +{{\cal O}}(\eta ^2).$
Of course, when computing matrix elements of these operators and then
forming bilinear combinations of the results to obtain the EM observables
terms of order $\eta ^2$ must be neglected if the operators themselves have
been expanded only to order $\eta $, since other terms will enter from
considering the neglected ${{\cal O}}(\eta ^2)$ contributions in Eqs.~(\ref
{sn29}--\ref{sn31}).

Note that we have only expanded in powers of $\eta $, having argued that
this is typically a small quantity. Indeed, if it is not small (for
instance, when considering very high momentum components in the nuclear wave
function, {\em viz.} $\eta \sim 1\leftrightarrow p\sim M$) then the
full expressions must be employed; however, in such circumstances the struck
nucleon is very far off-shell and the entire procedure becomes doubtful.
What we have not done is to expand in powers of the other dimensionless
momenta, $\kappa $, $\lambda $ or $\tau $. To see why this can be a problem
at moderate momentum transfers note that the typical observables that occur
(for example, $W_2=(G_E^2+\tau G_M^2)/(1+\tau)$) involve combinations of the
form factors where one often has $G_E^2$ together with $\left( \sqrt{\tau }%
G_M\right) ^2$. Using the dipole approximation to estimate the size of these
quantities we see that they become equal for protons at $\tau \cong
0.13\leftrightarrow |Q^2|\cong (670$ MeV/c$)^2$, for neutrons at $\tau
\cong 0.27\leftrightarrow |Q^2|\cong (980$ MeV/c$)^2$, or alternatively,
for isovector contributions at $\tau \cong 0.045\leftrightarrow
|Q^2|\cong (400$ MeV/c$)^2$ and for isoscalar contributions at $\tau
\cong 1.29\leftrightarrow |Q^2|\cong (2.1$ GeV/c$)^2$. Clearly for many
practical applications it is not valid only to retain terms of leading order
in $\sqrt{\tau }$ or $\kappa $. Moreover, as we see from the results presented
in Sect.~3.1, the distinction between $\kappa$ and $\sqrt{\tau}$ becomes
important for QE scattering at high momentum transfer.
In fact, as we see from the expressions
given above, it is unnecessary to make expansions in $\kappa $, $\lambda $
or $\tau $ at all. If one does, however, choose to do so, then at
intermediate momentum transfers the combination $G_M^{\prime }\equiv \sqrt{%
\tau }G_M$ should be regarded as being of leading order, and not of order $%
M^{-2}$ as is often assumed. With these caveats under some circumstances
Eq.~(\ref{sn31}) may be approximated by
\begin{equation}
{\bar{\bd J}}^{\bot }\cong -\frac{\sqrt{\tau }}\kappa \left\{ iG_M\left[
\bd{\kappa }\times \bd{\sigma }\right] -G_E\left[ \bd{\eta }%
-\left( \frac{\bd{\kappa }\cdot \bd{\eta }}{\kappa ^2}\right)
\bd{\kappa }\right] \right\} .  \label{sn32}
\end{equation}

At this point it should be mentioned that neglecting the second term
in Eq.~(\ref{sn29}) (the spin-orbit term), which is order $\kappa\eta$
compared with the first term which is of order unity, may not always be
justified. However, in spin-saturated systems such as the closed-shell
nuclei considered in the present work this term contributes only
quadratically and may safely be neglected when computing EM responses.
For instance, the contribution of the spin-orbit term to $R^L$ and
$R^L_{AV}$ in a spin-saturated system can be estimated using the RFG
model to be $\frac{3}{10}\eta_F^2 \frac{\tau}{1+\tau} \left(1+\frac{G_M^2}
{G_E^2} \right)$ and $\frac{3}{10}\eta_F^2 \frac{\tau}{1+\tau}
\left(1+\frac{G_M{\tilde G}_M}{G_E{\tilde G}_E} \right)$, respectively,
compared with the leading-order contributions $\frac{\kappa^2}{\tau} G_E^2$
and $\frac{\kappa^2}{\tau} G_E{\tilde G}_E$. For typical values of $\eta_F$
($\sim\frac{1}{4}$) the correction to $R^L$ is negligible ($<$ 4\%),
while the correction to $R^L_{AV}$ is sizable ($\sim$ 40\%) and has been
taken into account in the results presented in Sect.~3. There we see
that the latter response is quite small compared with ``normal'' response
functions and thus this correction is relatively unimportant.

As before, the vector NC analogs of these EM results are obtained simply by
making the replacements $G_{E,M}\rightarrow \tilde{G}_{E,M}$.
Likewise, the axial-vector results above may be expanded in powers
of $\eta $ following the procedures used for the EM case,
yielding the transverse axial-vector current operator to order $\eta $:
\begin{eqnarray}
{\bar{\bd J}}_5^{\bot } &=&\sqrt{1+\tau }{\tilde G}_A\Big\{ \bd{\sigma }%
^{\bot } \nonumber\\
&& \qquad \left. +\frac 1{2\left( 1+\tau \right) }
  \left( \left( \bd{\kappa }\cdot
\bd{\sigma }\right) \left[ \bd{\eta }-\left( \frac{\bd{\kappa }%
\cdot \bd{\eta }}{\kappa ^2}\right) \bd{\kappa }\right] -i\left(
\bd{\kappa }\times \bd{\eta }\right) \right) \right\}   \label{sn41}
\\
&\cong &\sqrt{1+\tau }{\tilde G}_A\bd{\sigma }^{\bot },  \label{sn42}
\end{eqnarray}
where in Eq.~(\ref{sn42}) the second term, involving corrections of order $%
\kappa \eta $ compared with the first term, has been neglected.

Finally, in the discussions in the main body of the paper we also refer
to the strict non-relativistic limit. In the present work we take this to
mean that the currents in Eqs.~(\ref{sn29},\ref{sn32},\ref{sn42})
are futher approximated by taking $\kappa$ and $\tau$ to be small compared
with unity to yield
\begin{eqnarray}
\bar J^0 &\longrightarrow& G_E \label{sn51}\\
{\bar{\bd J}}^{\bot } &\longrightarrow& -iG_M\left[
\bd{\kappa }\times \bd{\sigma }\right] +G_E\left[ \bd{\eta }%
-\left( \frac{\bd{\kappa }\cdot \bd{\eta }}{\kappa ^2}\right)
\bd{\kappa }\right]  \label{sn52}\\
{\bar{\bd J}}_5^{\bot } &\longrightarrow&
{\tilde G}_A\bd{\sigma }^{\bot }.  \label{sn53}
\end{eqnarray}


\subsection*{Appendix B: Reduced Matrix Elements in the Shell Model}

After a laborious but straightforward calculation
involving angular momentum algebra, the
different reduced matrix elements involved in the PV responses can be
written as follows: for the EM matrix elements one has
\beqa
\langle ph^{-1}, J\|M^J_{\rm em}\|0\rangle &=&
	\frac{(-1)^{j_p-1/2}}{\sqrt{4\pi}}\frac{\kappa}{\sqrt{\tau}}G_E
	\xi(\ell_p+\ell_h+J)\hat{j}_p\hat{j}_h\hat{J}
	\tresj{j_p}{j_h}{J}{1/2}{-1/2}{0}
\nonumber \\
&\times & \int_0^{\infty} dr r^2 j_J(qr) R_p(r) R_h(r)
\label{A1}
\eeqa

\beqa
\langle ph^{-1}, J\|T^{EJ}_{\rm em}\|0\rangle_{\rm mag} &=&
	\frac{(-1)^{j_p+1/2}}{\sqrt{4\pi J(J+1)}}
	\sqrt{\tau}G_M
	\xi(\ell_p+\ell_h+J)\hat{j}_p\hat{j}_h\hat{J}
	\tresj{j_p}{j_h}{J}{1/2}{-1/2}{0}
\nonumber \\
& \times & (\gamma_p-\gamma_h)
	\int_0^{\infty} dr r^2 j_J(qr) R_p(r) R_h(r)
\label{A2}
\eeqa

\beqa
\langle ph^{-1}, J\|T^{EJ}_{\rm em}\|0\rangle_{\rm conv} &=&
	\frac{(-1)^{j_p-1/2}}{\sqrt{4\pi J(J+1)}}
	\frac{\sqrt{\tau}}{\kappa}G_E
	\xi(\ell_p+\ell_h+J)\hat{j}_p\hat{j}_h\hat{J}
	\tresj{j_p}{j_h}{J}{1/2}{-1/2}{0}
\nonumber \\
&&\kern -1cm \times\frac{1}{2M}\int_0^{\infty} dr r j_J(qr)\frac{1}{q}
	\left\{\left[(\gamma_p-\gamma_h)(\gamma_p+\gamma_h+1)+J(J+1)\right]
	R_p(r)\frac{d}{dr}R_h(r) \right.
\nonumber \\
& + &\left.  \left[(\gamma_p-\gamma_h)
	(\gamma_p+\gamma_h+1)-J(J+1)\right]R_h(r)
	\frac{d}{dr}R_p(r)\right\}
\label{A3}
\eeqa

\beqa
& &\langle ph^{-1}, J\|T^{MJ}_{\rm em}\|0\rangle_{\rm mag} =
	\frac{i(-1)^{j_p-1/2}}{\sqrt{4\pi J(J+1)}}\sqrt{\tau}G_M
	\xi(\ell_p+\ell_h+J+1)\hat{j}_p\hat{j}_h\hat{J}
\nonumber \\
& & \times \tresj{j_p}{j_h}{J}{1/2}{-1/2}{0}
	\int_0^{\infty} dr r^2 j_J(qr)\frac{1}{q}
	\left[(\gamma_p+\gamma_h)\left(\frac{d}{dr}+
	\frac{1}{r}\right)+\frac{J(J+1)}{r}\right]
	R_p(r) R_h(r)
\nonumber \\
& & \label{A4}
\eeqa

\beqa
& &\langle ph^{-1}, J\|T^{MJ}_{\rm em}\|0\rangle_{\rm conv} =
	\frac{i(-1)^{j_p-1/2}}{\sqrt{4\pi J(J+1)}}\frac{\sqrt{\tau}}{\kappa}G_E
	\xi(\ell_p+\ell_h+J+1)\hat{j}_p\hat{j}_h\hat{J}
\nonumber \\
& & \times \tresj{j_p}{j_h}{J}{1/2}{-1/2}{0}
	\left[(\gamma_p+\gamma_h)(\gamma_p+\gamma_h+1)-J(J+1)\right]
	\frac{1}{2M}\int_0^{\infty} dr r j_J(qr) R_p(r) R_h(r),
\nonumber \\
&  & \label{A5}
\eeqa
together with their vector NC analogs, obtained by replacing $G_E$ with
$\tilde{G}_E$ and $G_M$ with $\tilde{G}_M$.

For the axial-vector NC operators, using the relations given by
Eqs.~(\ref{48},\ref{49}), one simply obtains

\begin{eqnarray}
 & &   \langle ph^{-1}, J\|\left(T^{EJ}_{\rm nc}\right)_{A}\|0\rangle
= \frac{i(-1)^{j_p-1/2}}{\sqrt{4\pi J(J+1)}}\sqrt{1+\tau}\tilde{G}_A
    \xi(\ell_p+\ell_h+J+1)\hat{j}_p\hat{j}_h\hat{J}
     \nonumber\\
& &\times    \tresj{j_p}{j_h}{J}{1/2}{-1/2}{0}
    \int_0^{\infty}dr\,r^2j_J(qr)\frac{1}{q}
    \left[(\gamma_p+\gamma_h)
           \left(\frac{d}{dr}+\frac{1}{r}\right)+
           \frac{J(J+1)}{r}
    \right] R_p(r)R_h(r) \nonumber\\
\label{A6}
\eeqa

\beqa
\langle ph^{-1}, J\|\left(T^{MJ}_{\rm nc}\right)_{A}\|0\rangle
&=& \frac{(-1)^{j_p+1/2}}{\sqrt{4\pi J(J+1)}}\sqrt{1+\tau}\tilde{G}_A
    \xi(\ell_p+\ell_h+J)\hat{j}_p\hat{j}_h\hat{J}
    \tresj{j_p}{j_h}{J}{1/2}{-1/2}{0}
     \nonumber\\
&\times & \left(\gamma_p-\gamma_h \right)
	\int_0^{\infty}dr\,r^2j_J(qr) R_p(r)R_h(r).
\label{A7}
\end{eqnarray}
In all of the above equations we have used the sign function $\xi$ defined as
\beq
\xi(\ell_p+\ell_j+J)\equiv
	\frac{1+(-1)^{\ell_p+\ell_h+J}}{2}.
\label{A8}
\eeq
The functions,
$R_p$ ($R_h$) represent the radial single-particle (single-hole)
wave functions, and we have introduced the term
$\gamma_i\equiv (\ell_i-j_i)(2j_i+1)$.

\newpage

\subsection*{Appendix C: Fermi Gas Model}

In this appendix we summarize the main features of the relativistic
Fermi gas model (RFG), drawing on the detailed discussions in
Refs.~\cite{Don92,Alb88,Alb90,Alb93,Bar94}.
The purpose in presenting this material again is twofold: first,
we make comparisons with the RFG and its extensions~\cite{Bar94} when
discussing PV electron scattering in the results section of the present
paper; secondly, there are approximations that we make when employing
the CSM and the RFG approach provides the basic motivation for these, as
discussed below.

We begin by re-stating the procedures followed in deriving the RFG
responses in the above-cited work. The exact, fully-relativistic,
on-shell, single-nucleon current matrix elements obtained by inserting
spin spinors before and after the current operators given in
Eqs.~(\ref{sn12}--\ref{sn17}) and (\ref{sn35}--\ref{sn37}) may be computed
in a straightforward way. The results are given in Ref.~\cite{Alb90}
(in fact, multiplied by factors that occur in the invariant phase-space
factors). These are the matrix elements for elastic electroweak
scattering from on-shell, moving nucleons. Taking, as usual, the RFG
step-function momentum distribution characterized by dimensionless Fermi
momentum $\eta_F\equiv p_F/M$ and integrating over all nucleons in the
filled Fermi sea, one obtains the familiar response functions
\begin{eqnarray}
R^{L,T} &=& R_0 (\kappa,\lambda) U^{L,T} (\kappa,\lambda) \label{fg1}\\
R^{L,T}_{AV} &=& R_0 (\kappa,\lambda) \tilde{U}^{L,T} (\kappa,\lambda)
  \label{fg2}\\
R^{T'}_{VA} &=& R_0 (\kappa,\lambda) \tilde{U}^{T'} (\kappa,\lambda),
  \label{fg3}
\end{eqnarray}
where the dimensionless momentum and energy transfers are defined
in Appendix~A. One may consider both the Pauli-blocked and
non-Pauli-blocked regimes ($\kappa<\eta_F$ and $\kappa>\eta_F$, respectively).
Focusing here on the latter and following
Ref.~\cite{Alb88} we define a scaling variable
\begin{eqnarray}
\psi &\equiv& \left[ \frac{1}{\xi_F}\left( \kappa\sqrt{1+\frac{1}{\tau}}
   -(1+\lambda) \right)\right]^{1/2} \nonumber\\
&&\qquad\qquad \times\left(\theta[\lambda-\lambda_0] -\theta[\lambda_0-\lambda]
   \right), \label{fg4}
\end{eqnarray}
where $\xi_F\equiv\sqrt{1+\eta_F^2}-1\cong \eta_F^2/2$, since $\eta_F$ is
typically small, as discussed in Appendix~A. The RFG QE peak
occurs at $\lambda=\lambda_0\equiv \frac{1}{2}[\sqrt{1+4\kappa^2}-1]$.

In the non-Pauli-blocked regime the overall responses are characterized by
\begin{equation}
R_0 (\kappa,\lambda) = \frac{3{\cal N}\xi_F}{4M\kappa\eta_F^3}
  (1-\psi^2)\theta(1-\psi^2), \label{fg5}
\end{equation}
where ${\cal N}$ is $Z$ ($N$) when protons (neutrons) are involved (the
expressions for the Pauli-blocked regime are discussed in Ref.~\cite{Alb88}
--- here our main emphasis is on the non-Pauli-blocked region although
in Sect.~3 some results are also presented for $\kappa<\eta_F$). The
dependences on the single-nucleon content in the problem are then
isolated in the factors
\begin{eqnarray}
U^L (\kappa,\lambda) &=& \frac{\kappa^2}{\tau}\left\{ \left[
   (1+\tau)W_2(\tau)-W_1(\tau) \right] + W_2(\tau)\Delta \right\} \label{fg6}\\
U^T (\kappa,\lambda) &=& \frac{\kappa^2}{\tau}\left\{ \left[
   (1+\tau)\tilde{W}_2(\tau)-\tilde{W}_1(\tau) \right] +
   \tilde{W}_2(\tau)\Delta \right\}          \label{fg7}\\
\tilde{U}^L (\kappa,\lambda) &=& 2W_1(\tau)+W_2(\tau)\Delta \label{fg8}\\
\tilde{U}^T (\kappa,\lambda) &=& 2\tilde{W}_1(\tau)+\tilde{W}_2(\tau)\Delta
         \label{fg9}\\
\tilde{U}^{T'} (\kappa,\lambda) &=& \sqrt{\tau(1+\tau)}\tilde{W}_3(\tau)
   \left\{1+\tilde{\Delta}\right\}.         \label{fg10}
\end{eqnarray}
Here the nucleon form factors are those discussed in the main body of the
text (for protons or neutrons, as is appropriate). The factors $\Delta$ and
$\tilde{\Delta}$ were introduced in Refs.~\cite{Don92,Alb88}:
\begin{eqnarray}
\Delta &=& \frac{\tau}{\kappa^2}\xi_F (1-\psi^2)\left\{
   \kappa\sqrt{1+\frac{1}{\tau}} + \frac{1}{3}\xi_F (1-\psi^2) \right\}
   \label{fg11}\\
\tilde{\Delta} &=& \frac{1}{\kappa} \sqrt{ \frac{\tau}{1+\tau} } \left\{
   \frac{1}{2} \xi_F (1-\psi^2) \right\}. \label{fg12}
\end{eqnarray}
Clearly these two terms provide relatively small contributions to the
totals in Eqs.~(\ref{fg6}--\ref{fg10}), since they both involve the
factor $\xi_F$ which is very small. These are contributions of order
$\eta_F^2$ and typically enter at the few percent level in the kinematic
regime of interest here. Indeed, in Appendix~A we argued that such
${\cal O}(\eta^2)$ effects could be neglected --- we have done so in
the treatment of the CSM in the present work. In Sect.~3.1 we present
results for the RFG (and these agree with the past results cited above)
both with the exact relativistic current matrix elements and with the
approximations discussed in Appendix~A. Moreover, we also display in
Sect.~3.1 some results for what we defined in Appendix~A to be the strict
non-relativistic approximation (see Eqs.~(\ref{sn51}--\ref{sn53})) to
illustrate how poor those commonly-made approximations can be (see
Fig.~2).

A second point to be drawn from this summary of the RFG formalism, in
addition to bringing out the roles played by the single-nucleon current matrix
elements, is seen by returning to $R_0$ defined in Eq.~(\ref{fg5}). The
basic RFG behavior in the non-Pauli-blocked region when the single-nucleon
factors (the $U$'s and $\tilde{U}$'s) are removed is parabolic in the
scaling variable $\psi$. For the RFG this quantity ranges between $-1$ and
$+1$, with the QE peak occurring when $\psi=0$. An excellent
approximation to the scaling variable may be shown to be
(see Ref.~\cite{Alb93})
\begin{equation}
\psi \cong \frac{1}{\eta_F} \left[ \frac{\lambda (1+\lambda)}{\kappa}
   -\kappa \right], \label{fg13}
\end{equation}
making it easy to see that the QE peak ($\psi=0$) corresponds
to $\lambda (1+\lambda)=\kappa^2 \leftrightarrow \lambda=\tau
\leftrightarrow \omega = |Q^2|/2M$, the relativistic kinematical
relation. The non-relativistic version of Eq.~(\ref{fg13}) is
\begin{equation}
\psi_{nr} \cong \frac{1}{\eta_F} \left[ \frac{\lambda}{\kappa}
   -\kappa \right] \label{fg14}
\end{equation}
and yields the non-relativistic relation $\lambda=\kappa^2 \leftrightarrow
\omega=q^2/2M$ for the peak position. In the past Fermi gas work cited above
it was shown that a very good approximation is to employ the non-relativistic
many-body responses (of course, using good single-nucleon currents), but
with the replacement $\lambda\rightarrow \lambda(1+\lambda)$
or, for dimensionful variables, $\omega\rightarrow\omega(1+\omega/2M)$. The
replacement is made only in the many-body aspects of the problem, not in
the single-nucleon form factors which are evaluated at the correct value
of $\tau$, namely that given by the electron kinematic variables.
In Sect.~3.1 we present results to illustrate the quality of this
replacement in the nuclear kinematics. With this as motivation, we follow the
same ``relativizing'' procedure for the kinematics of the CSM as well by
using the relativistic relation in the energy-conserving $\delta$-function
by making the same replacement.





\newpage

\begin{table}[ph]
\begin{center}
\begin{tabular}{rrrrrrrr}
 & &
\multicolumn{1}{c}{$V_0$[MeV]} &
\multicolumn{1}{c}{$V_{ls}$[MeV]} &
\multicolumn{1}{c}{$R_0$[fm]} &
\multicolumn{1}{c}{$a_0$[fm]} &
\multicolumn{1}{c}{$R_{ls}$[fm]} &
\multicolumn{1}{c}{$a_{ls}$[fm]} \\\hline
$^{16}$O  & p & 52.5 &  7.00 & 3.20 & 0.53 & 3.20 & 0.53 \\
          & n & 52.5 &  6.54 & 3.20 & 0.53 & 3.20 & 0.53 \\
$^{40}$Ca  & p & 57.5 & 11.11 & 4.10 & 0.53 & 4.10 & 0.53 \\
          & n & 55.0 &  8.50 & 4.10 & 0.53 & 4.10 & 0.53 \\
$^{208}$Pb & p & 60.4 &  6.75 & 7.46 & 0.79 & 7.20 & 0.59 \\
          & n & 44.3 &  6.08 & 7.46 & 0.66 & 6.96 & 0.64 \\ \hline
\end{tabular}
\end{center}
\caption{Parameters of the Woods-Saxon potential.
They have been taken from Refs.~\protect\cite{Ama94b} for $^{16}$O,
\protect\cite{Ama94c} for $^{40}$Ca, and \protect\cite{Rin78} for $^{208}$Pb.}
\end{table}

\begin{table}[ph]
\begin{center}
\begin{tabular}{cccc}
Nucleus    & $q$[MeV/c] & $J_{max}$ & $J_{con}$ \\ \hline\hline
$^{16}$O   & 300 & 15        &  8        \\
           & 500 & 15        & 12        \\
           & 700 & 20        & 20        \\ \hline
$^{40}$Ca  & 300 & 15        & 11        \\
           & 500 & 15        & 15        \\
           & 700 & 23        & 23        \\ \hline
$^{208}$Pb & 300 & 20        & 18        \\
           & 500 & 23        & 23        \\
           & 700 & 30        & 30        \\ \hline
\end{tabular}
\end{center}
\caption{Multipoles used in the calculation.}
\end{table}

\newpage




\begin{center}
{\bf FIGURE CAPTIONS}
\end{center}
\begin{enumerate}

\item One boson exchange diagrams considered in this work.

\item Electromagnetic response functions $R^L$ and $R^T$ for $^{40}$Ca
      shown at $q=$ 500, 700 and 1000 MeV/c: (solid curves) full RFG model;
      using the approximate currents given in Sect.~2.2, (long-dashed curves)
      with the replacement $\lambda\rightarrow \lambda(1+\lambda)$ made to the
      non-relativistic FG model as discussed in Appendix~C or
      (dot-dashed curves) without this replacement; as in the previous case,
      but using the non-relativistic currents obtained by setting the
      factor $\kappa/\sqrt{\tau}\rightarrow 1$ in Eqs.~(\ref{20}) and
      (\ref{23}) with (short-dashed curves) or
      without (dotted curves) the $\lambda\rightarrow \lambda(1+\lambda)$
      replacement. In all cases $p_F$ is fixed at 235 MeV/c.

\item Electromagnetic response functions $R^L$ and $R^T$ for $^{16}$O at
      $q=$ 300, 500 and 700 MeV/c: (solid curves) CSM; (other curves)
      full RFG model, (dashed curves, $p_F=$ 215 MeV/c) and
      (dotted curves, $p_F=$ 225 MeV/c).

\item Electromagnetic response function $R^L$ for $^{40}$Ca and $^{208}$Pb at
      $q=$ 300, 500 and 700 MeV/c. The labeling in this and most of the
      following figures is: (solid curves) CSM; (other curves)
      full RFG model. The latter are the following: for $^{40}$Ca two are
      shown, (dashed curves, $p_F=$ 235 MeV/c) and (dotted curves, 245 MeV/c);
      for $^{208}$Pb three are shown, (long-dashed curves,
      $p_{Fp}=$ 250 MeV/c, $p_{Fn}=$ 288 MeV/c), (short-dashed curves,
      $p_{Fp}=p_{Fn}=$ 250 MeV/c) and (dotted curves,
      $p_{Fp}=$ 230 MeV/c, $p_{Fn}=$ 265 MeV/c), as discussed in the
      text. In fact, for the longitudinal response the short- and
      long-dashed curves essentially coincide.

\item As in Fig.~4, except now for the EM response $R^T$.

\item As in Fig.~4, except now for the PV response $R^T_{AV}$.

\item As in Fig.~4, except now for the PV response $R^{T'}_{VA}$.

\item As in Fig.~4, except now for the PV response $R^L_{AV}$.

\item Parity-violating asymmetry for $^{16}$O, $^{40}$Ca and $^{208}$Pb at
      $q=$ 300, 500 and 700 MeV/c and at $\theta_e=$ 10$^{\circ}$ and
      150$^{\circ}$ (labeling of curves as in Fig.~4). The difficulty in
      distinguishing the curves shows the nuclear model insensitivity of
      the results.

\item The sum-rule asymmetry (SRA) defined in Eq.~(\ref{r1}) shown
      for $^{40}$Ca and $^{208}$Pb at $q=$ 300, 500 and 700 MeV/c as a
      function of scattering angle $\theta_e$ (labeling of curves as in
Fig.~4).
      The difficulty in distinguishing the curves, especially in the case of
      $^{40}$Ca, reflects the high degree of nuclear model insensitivity for
      this observable.

\item The sum-rule asymmetry (SRA) defined in Eq.~(\ref{r1}) shown
      for $^{16}$O (solid curves), $^{40}$Ca (dashed curves) and $^{208}$Pb
      (dotted curves) at $q=$ 300, 500 and 700 MeV/c as a function of
      scattering angle $\theta_e$. The CSM is used in all three cases.
      The difficulty in distinguishing the curves (the results for $^{16}$O
      and $^{40}$Ca are essentially equal) indicates the universality of
      this observable for a wide range of nuclei.

\item As in Fig.~10, but now showing families of curves with magnetic
      strangeness (with $\mu_s=-0.35$) and without magnetic strangeness
      (indicated with a ``0'' in the case of $^{40}$Ca where the families
      of curves can be distinguished at high $q$). In this case the curves
      are seen to
      coalesce into sets determined by the value of $\mu_s$ and
      not by the nuclear model variations, as discussed in more depth in
      the text. The families of curves are labeled as in Fig.~4.

\item As in Fig.~10, but now focusing on scattering at backward angles and
      showing families of curves with different
      values of the isovector, axial-vector form factor of the nucleon.
      The curves marked with a $(\pm)$ correspond to
      setting $g_A=1.262\pm$10\%, while the remaining curves
      have this quantity set to 1.262. The families of
      curves are labeled as in Fig.~4. At intermediate-to-high values of
      $q$ the curves are seen to group into sets determined by the
      axial-vector coupling and not so much by the nuclear model
      variations.

\item As in Fig.~10, but now focusing on scattering at forward angles and
      showing families of curves with different
      values of the neutron electric form factor, the standard Galster
      parameterization with $\lambda_n=5.6$ (see Ref.~\cite{Mus94}) and
      values of this parameter that yield $\pm$ 10\% variation in $G_E^n$
      at kinematics where the form factor peaks, namely $\lambda_n=4.2$
      and 7.3 (marked $(\pm)$ in the figure when the groups are
      distinguishable).
      The families of curves are labeled as in Fig.~4. The nuclear model
      dependences and the differences due to variations in $G_E^n$ at
      this level would clearly be quite difficult to separate.

\item As in Fig.~14, but now showing families of curves with different
      parameterizations for $G_E^{(s)}$: the curves marked (I) have no
      $G_E^{(s)}$ contributions, marked (II) have $\rho_s=-3$ and
      $\lambda_E^{(s)}=5.6$ (``modest electric strangeness''),
      while those marked (III) have $\rho_s=-3$ and $\lambda_E^{(s)}=0$
      (``large electric strangeness''). In all cases the magnetic strangeness
      form factor has been taken to be the standard parameterization
      discussed in the text. Clearly for high $q$ and forward
      angles the nuclear model dependences are considerably weaker than
      the dependence on the parameterization of electric strangeness in
      the nucleon.

\end{enumerate}

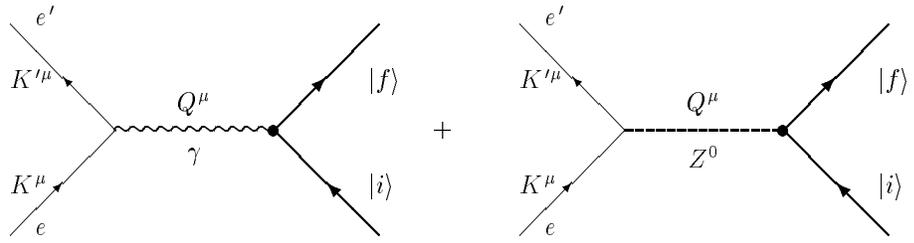
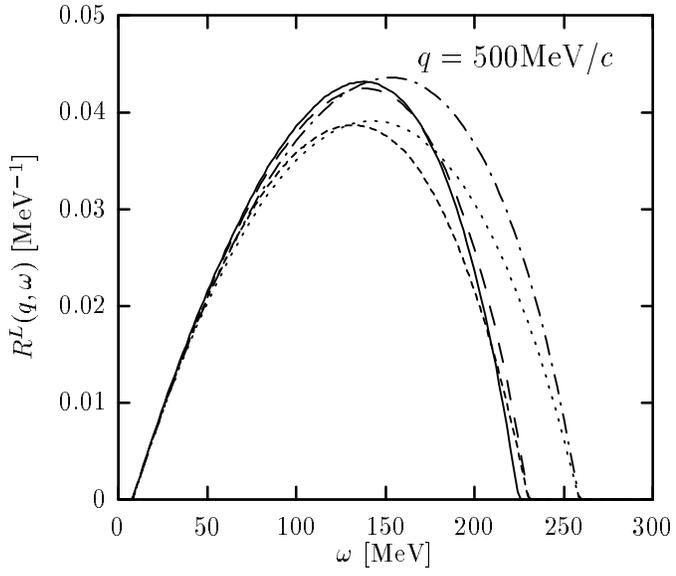
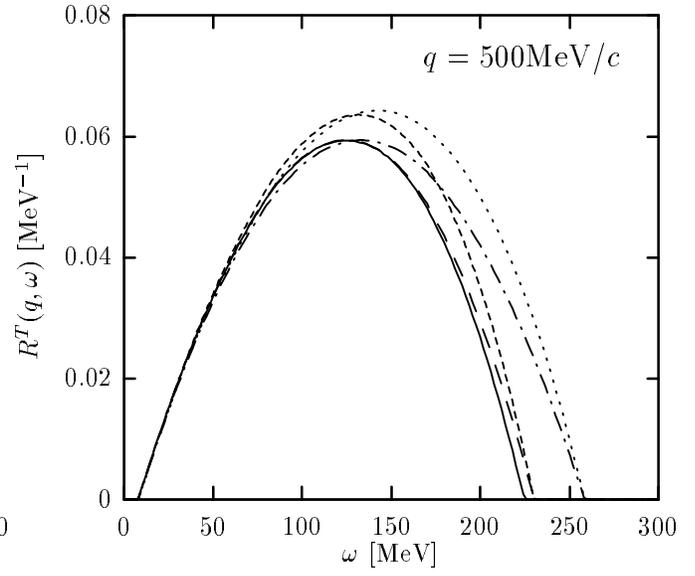
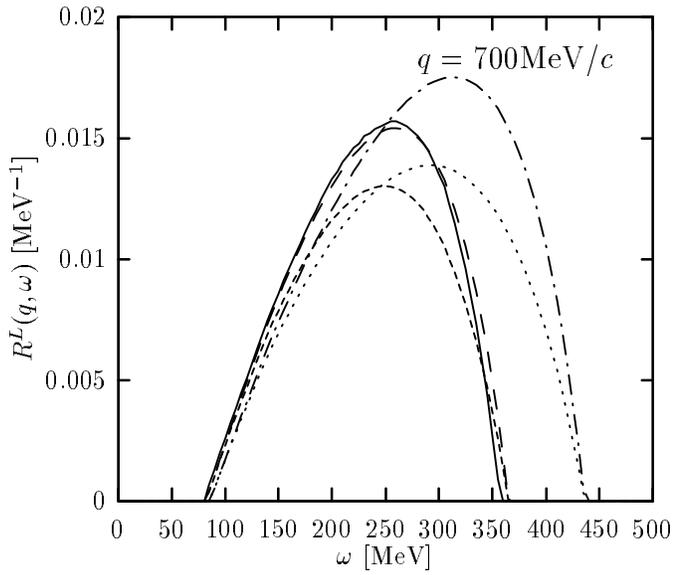
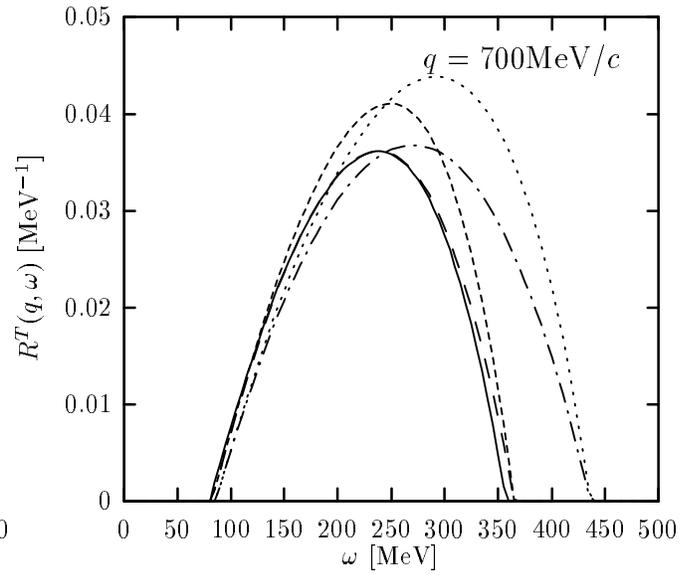
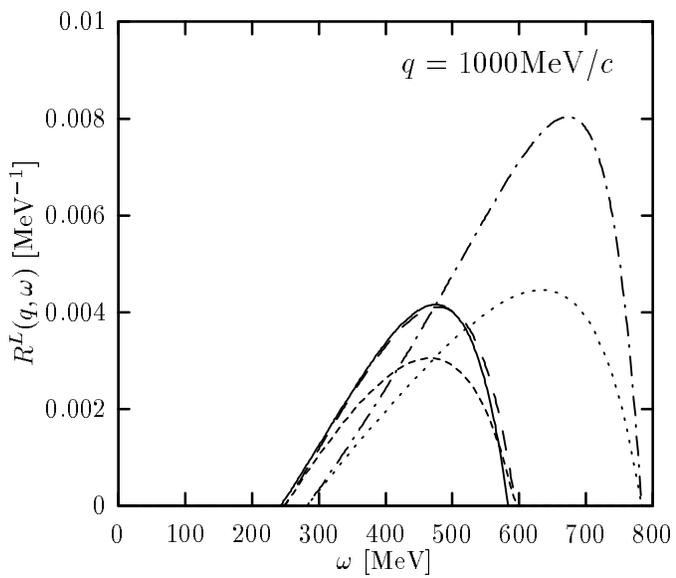
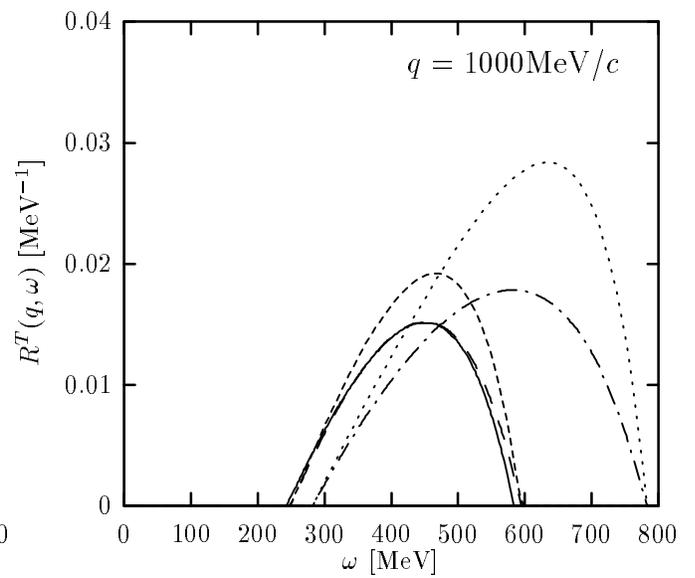
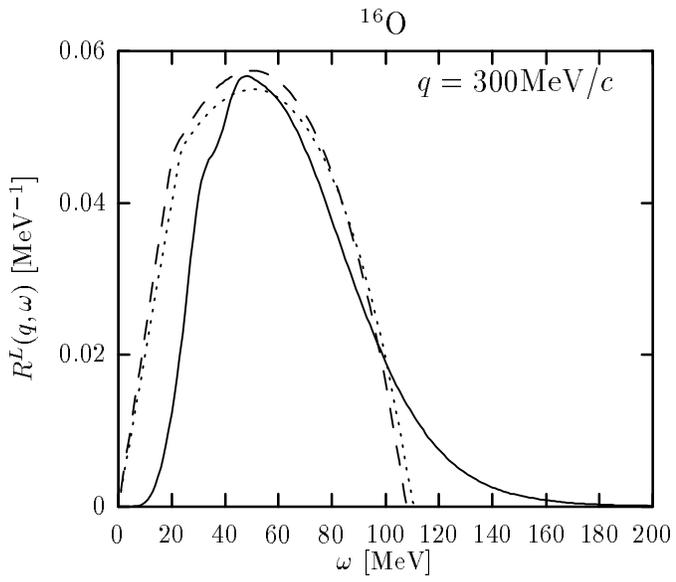
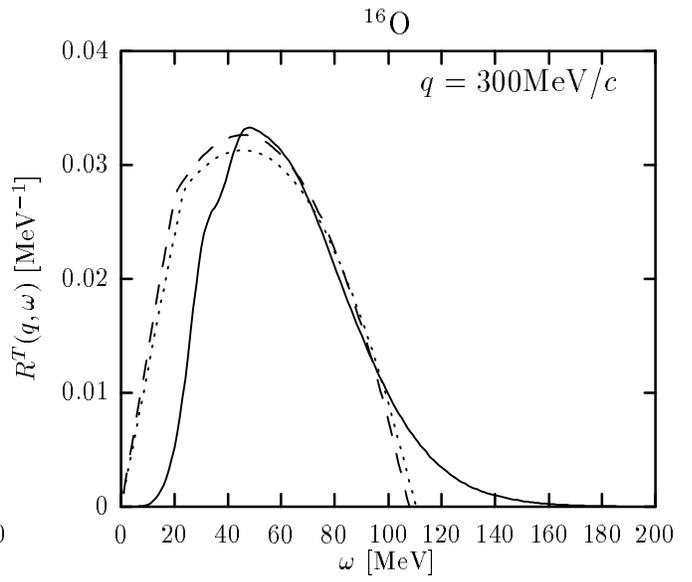
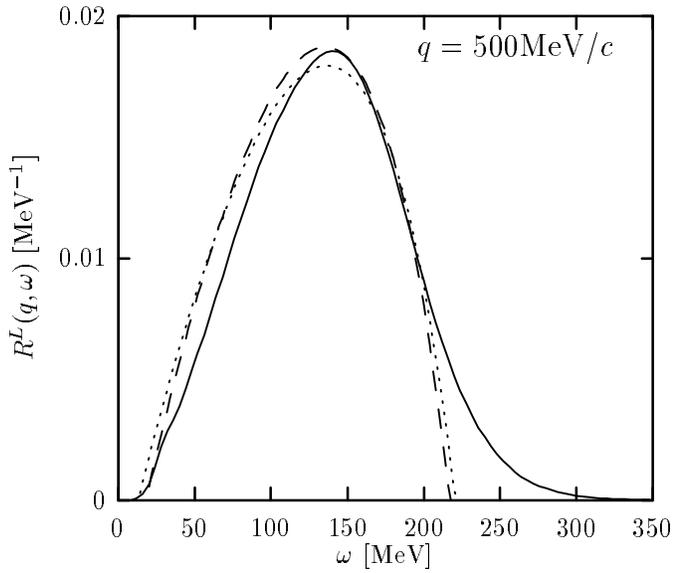
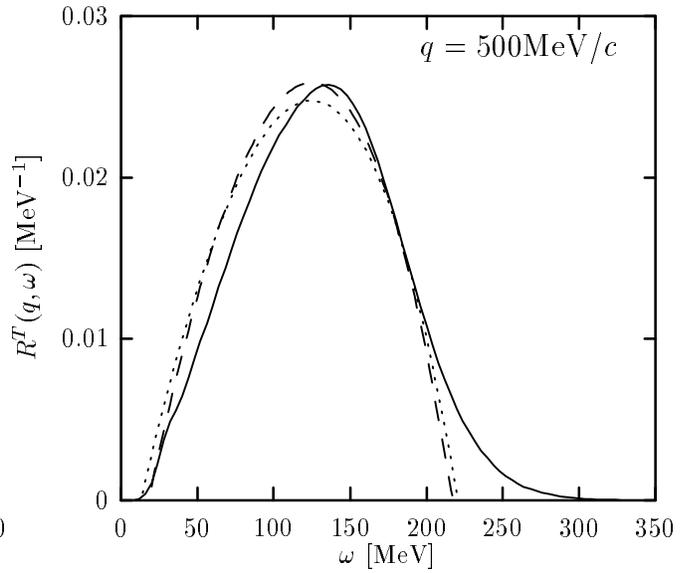
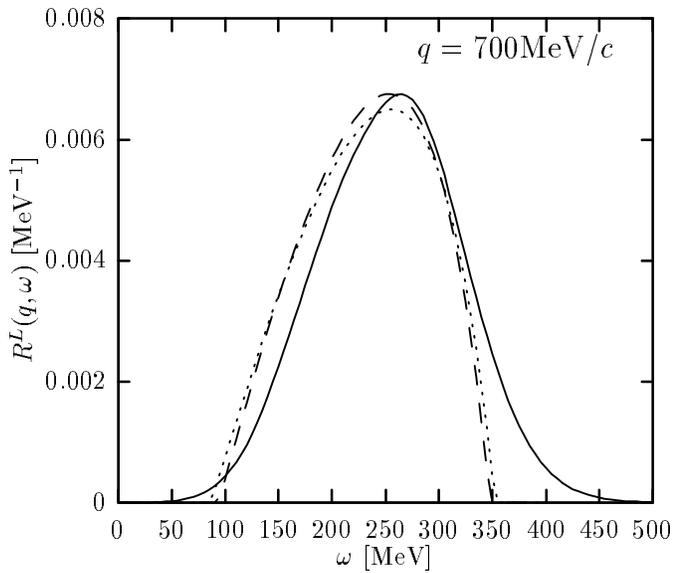
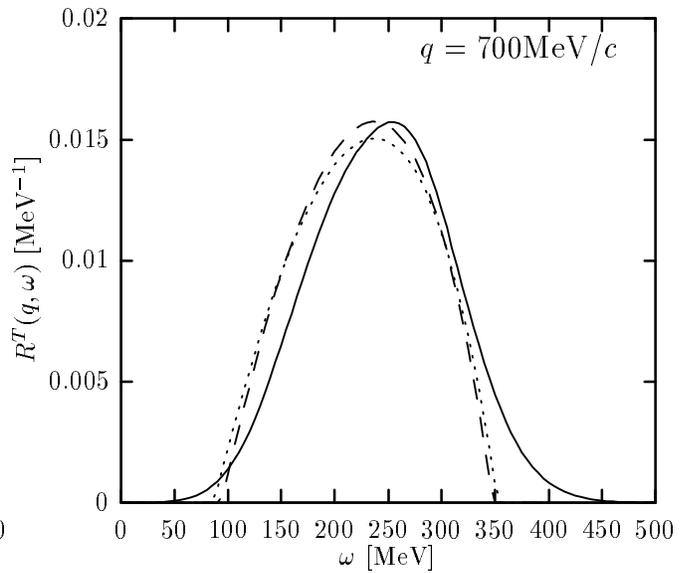
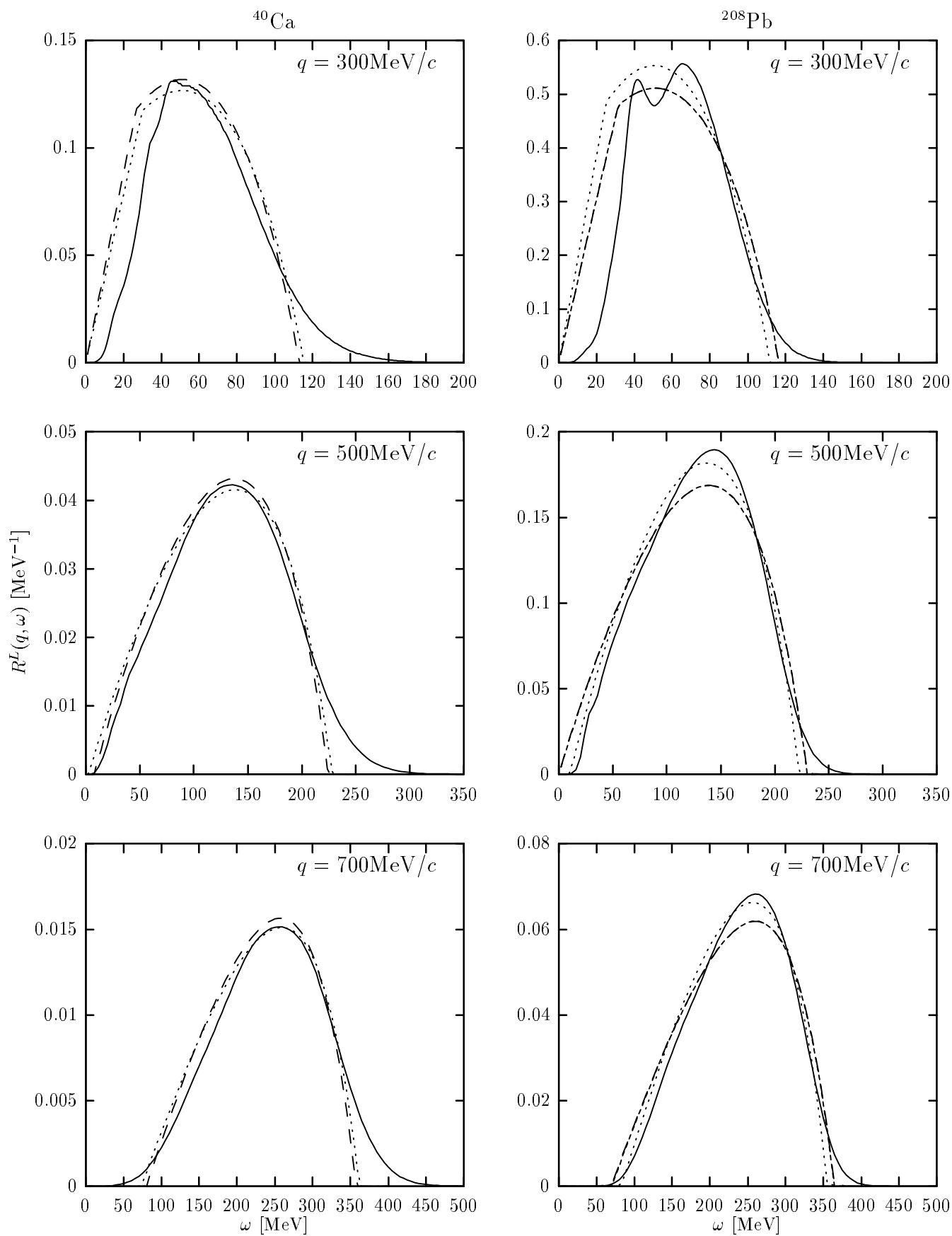
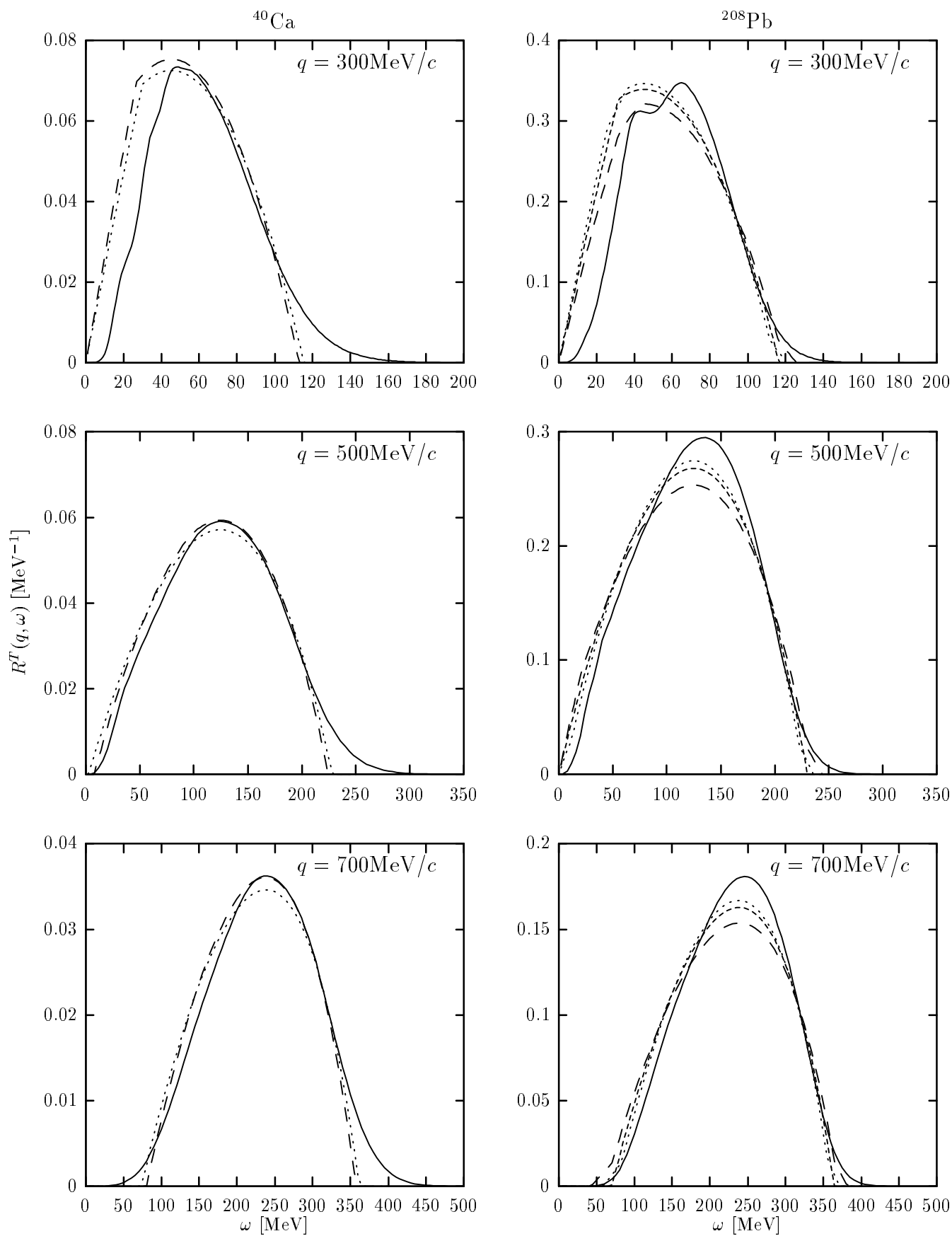
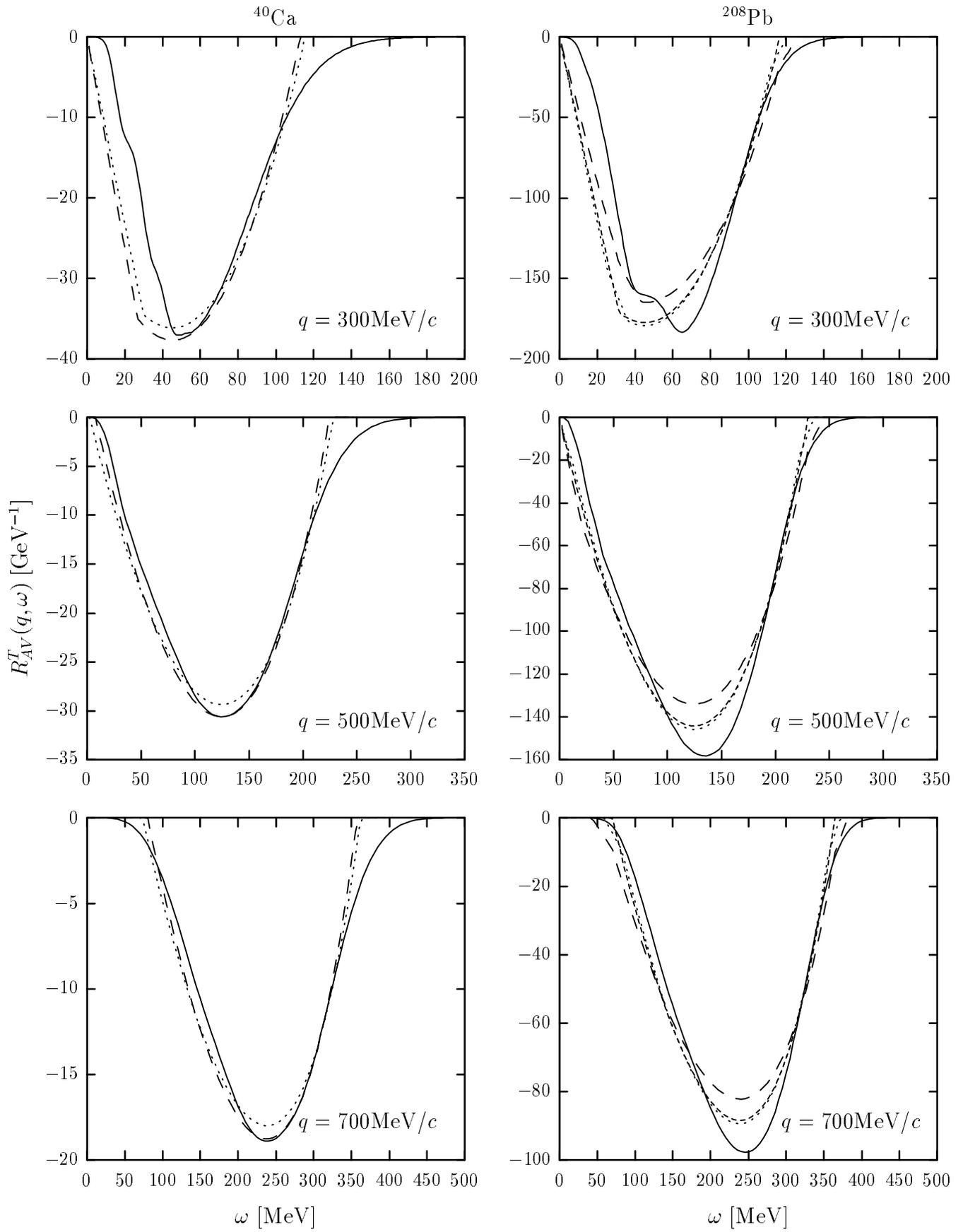
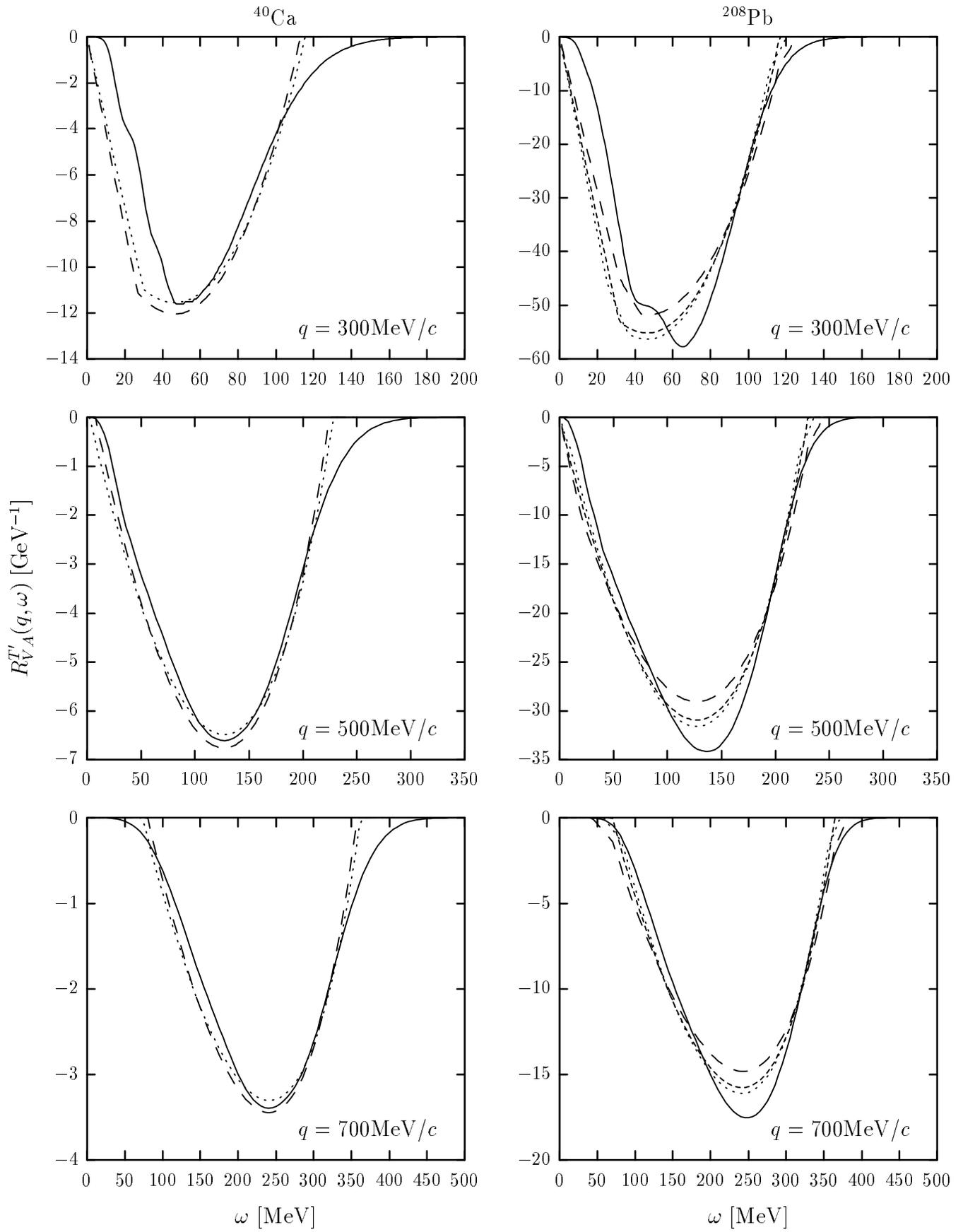
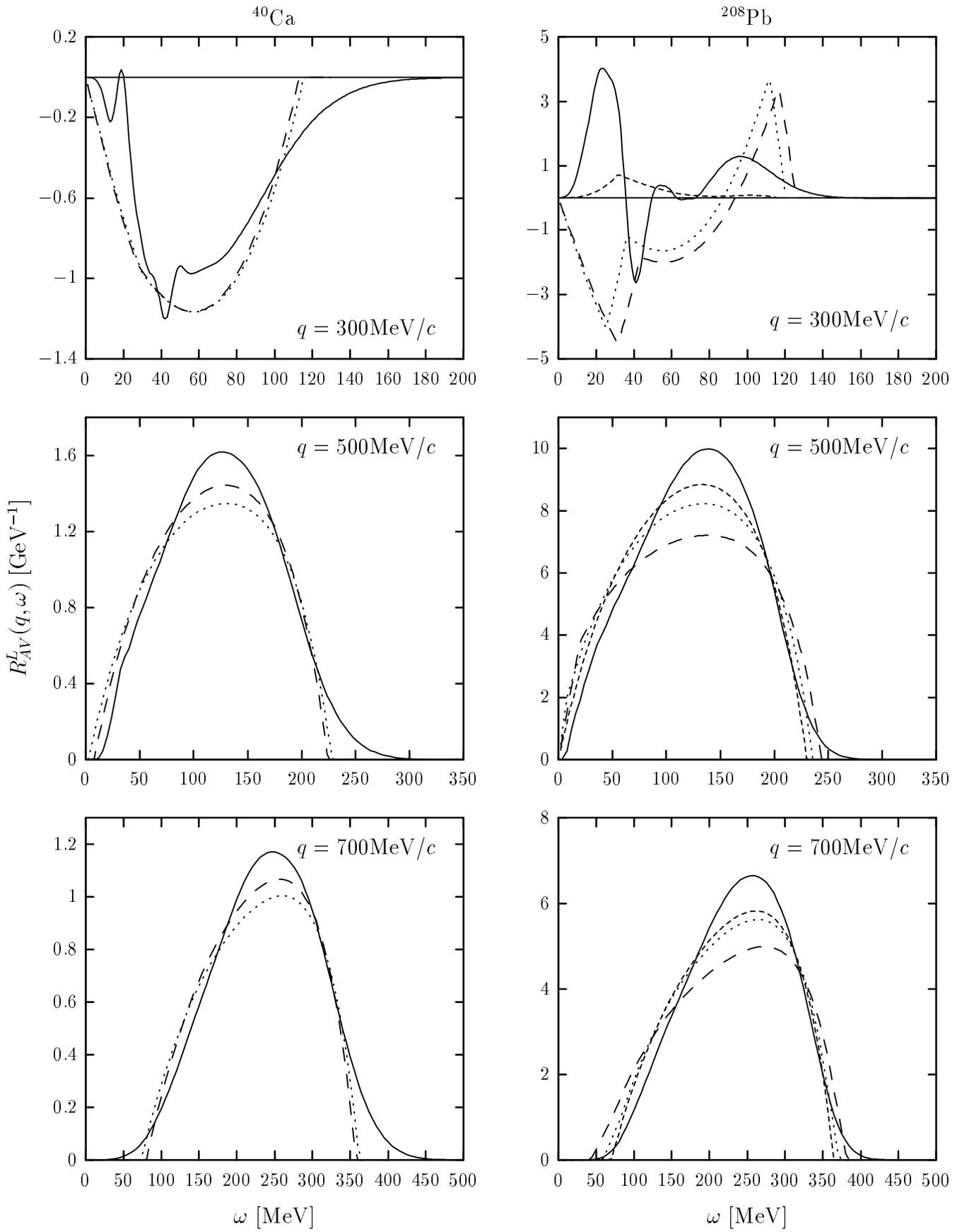
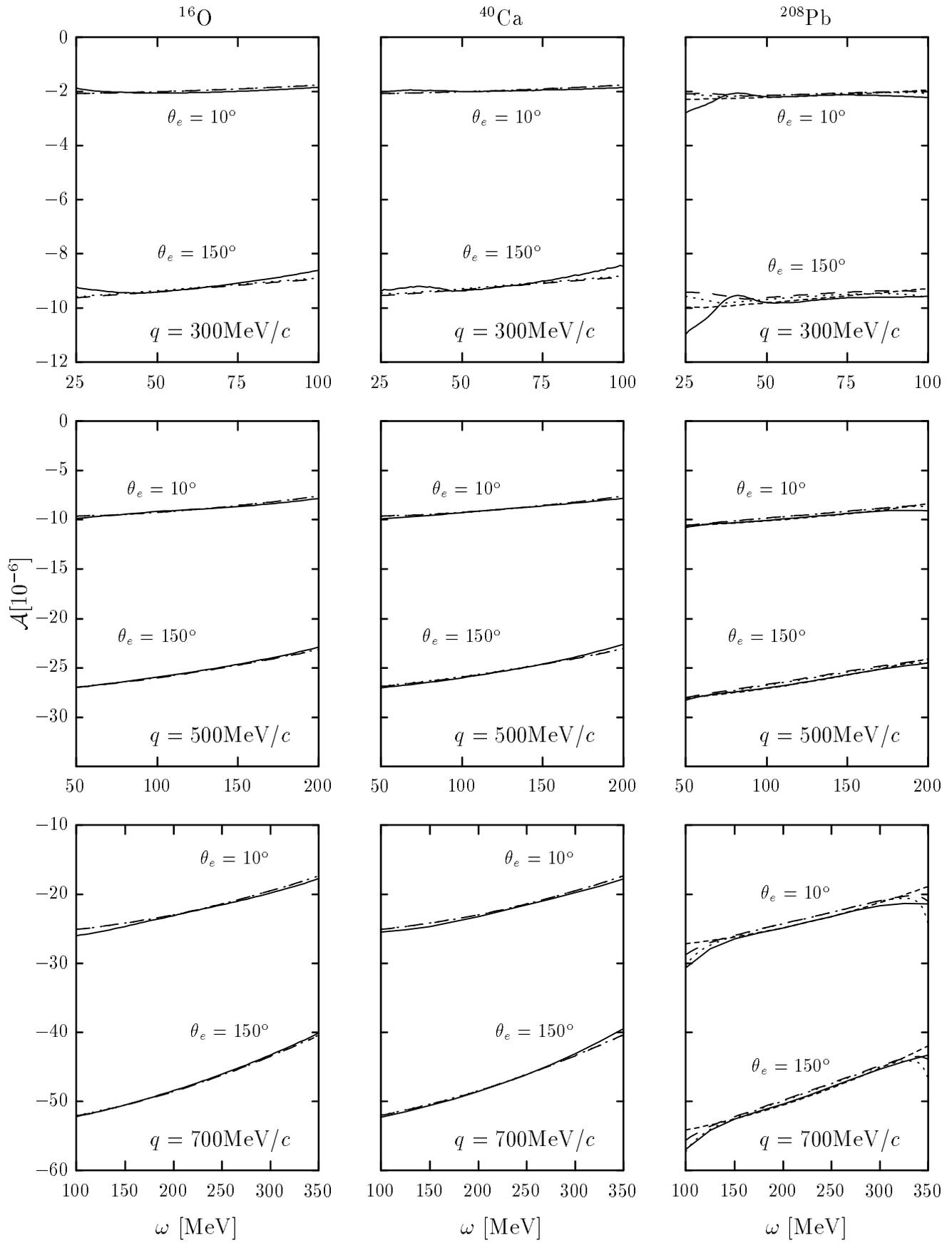
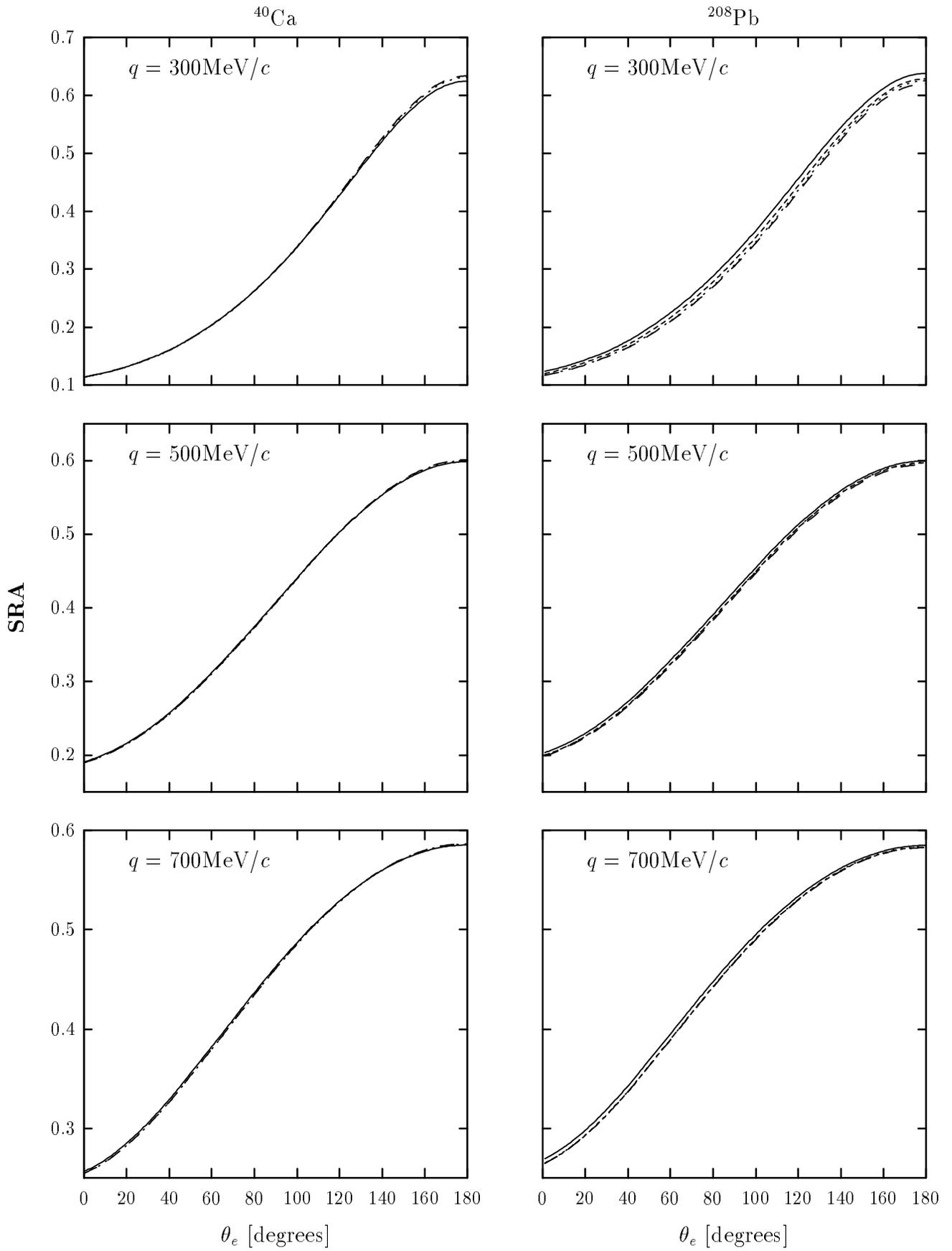
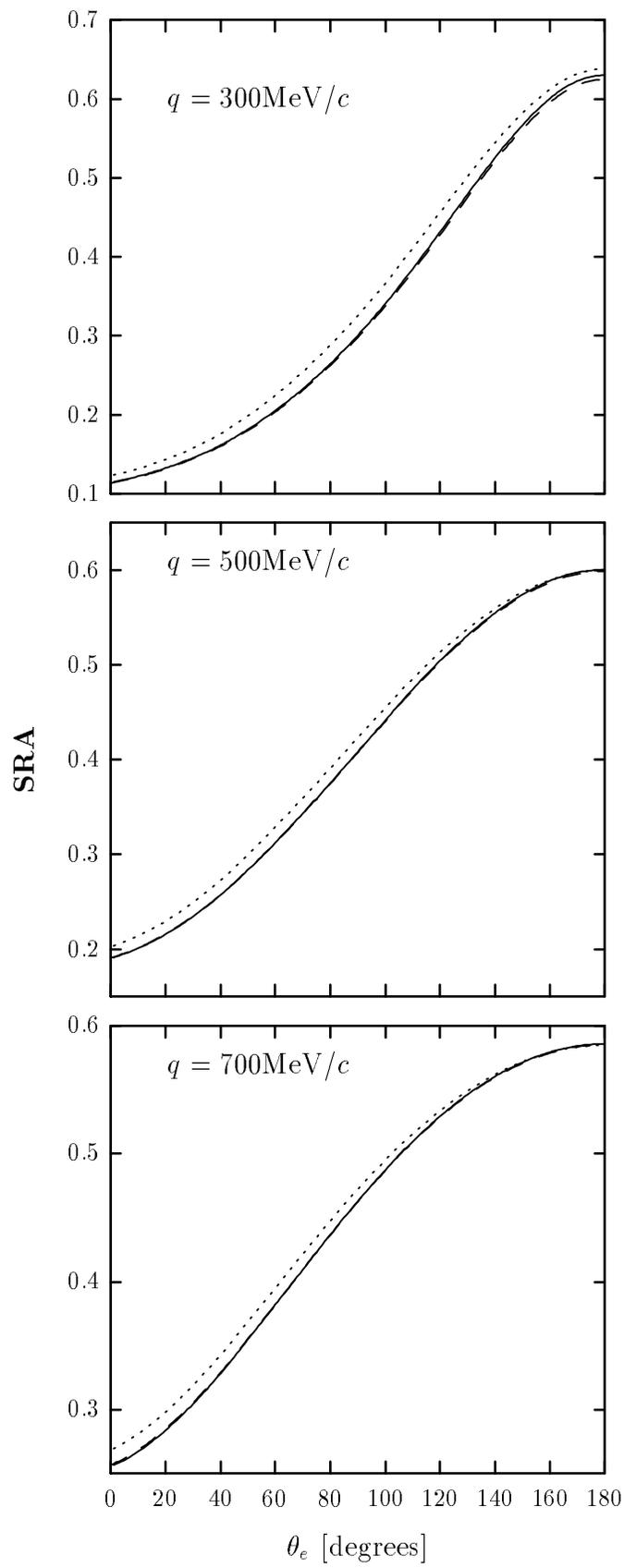
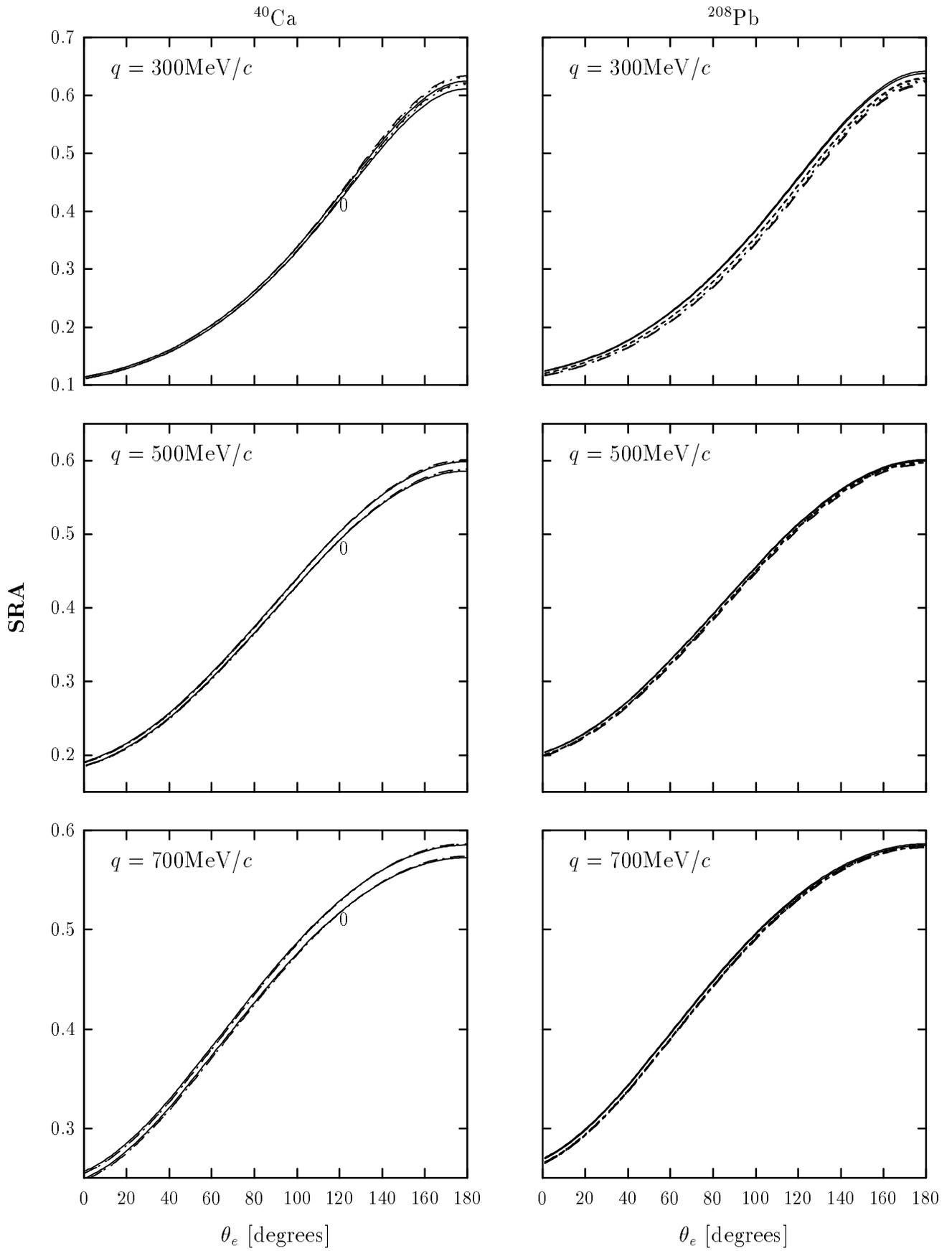
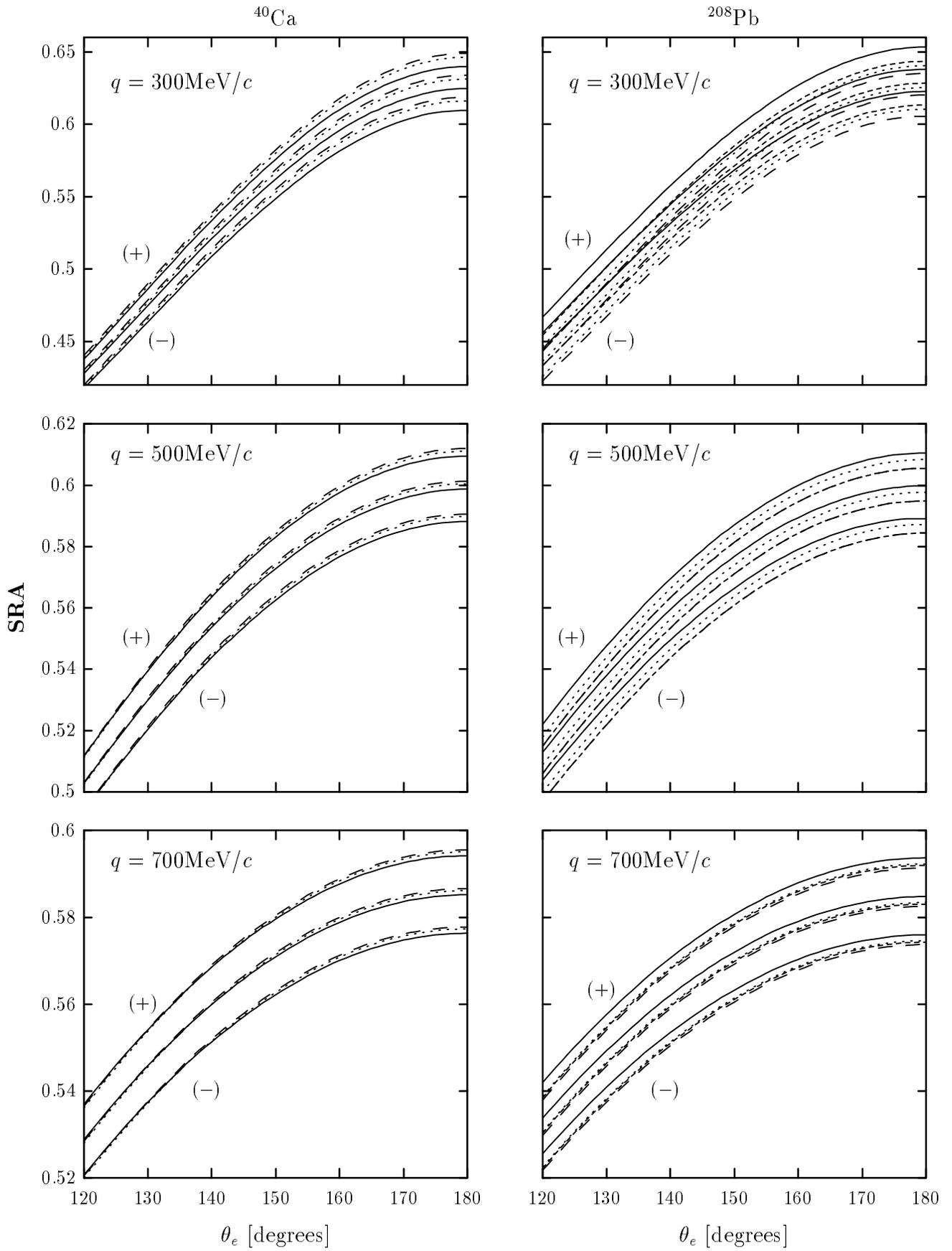
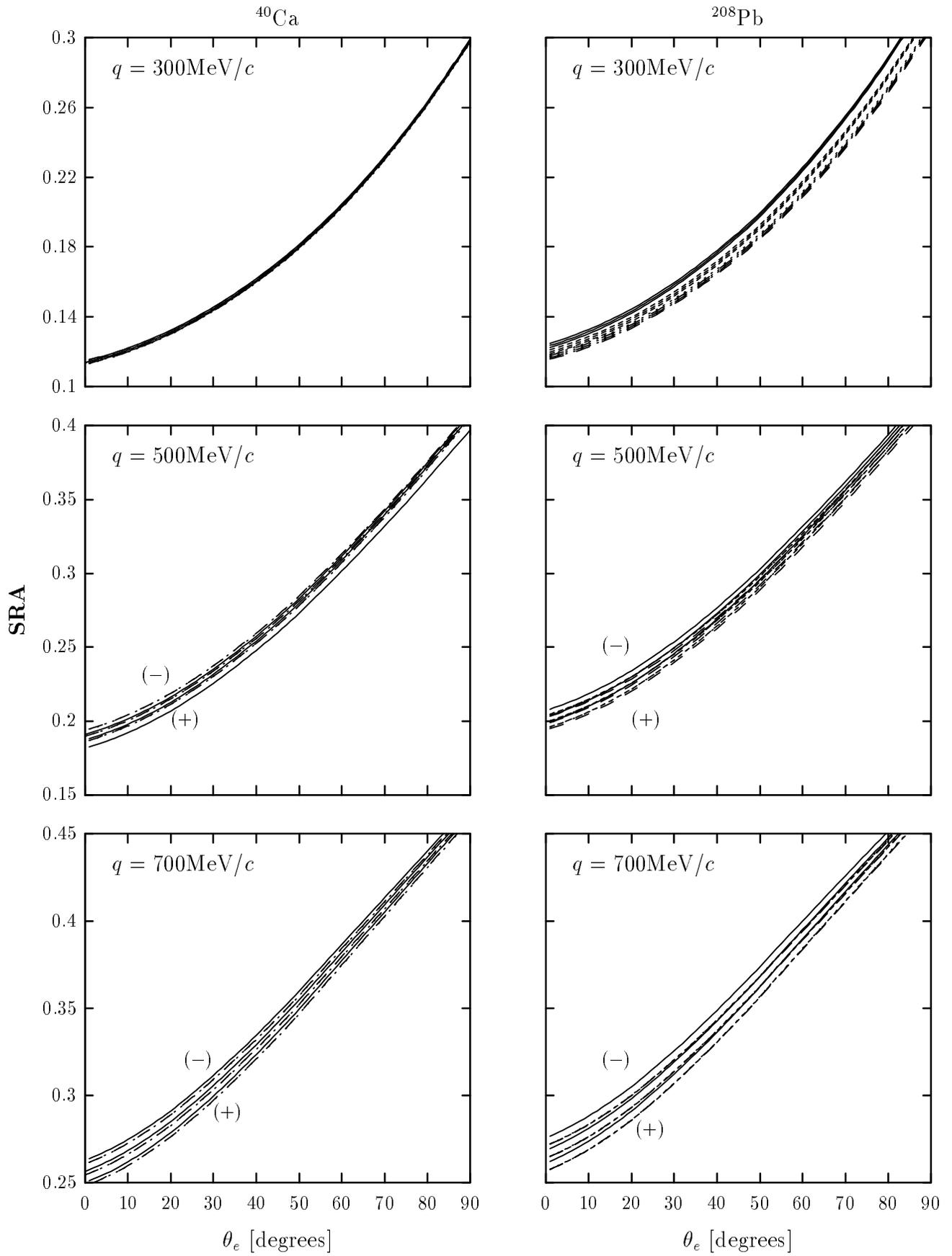
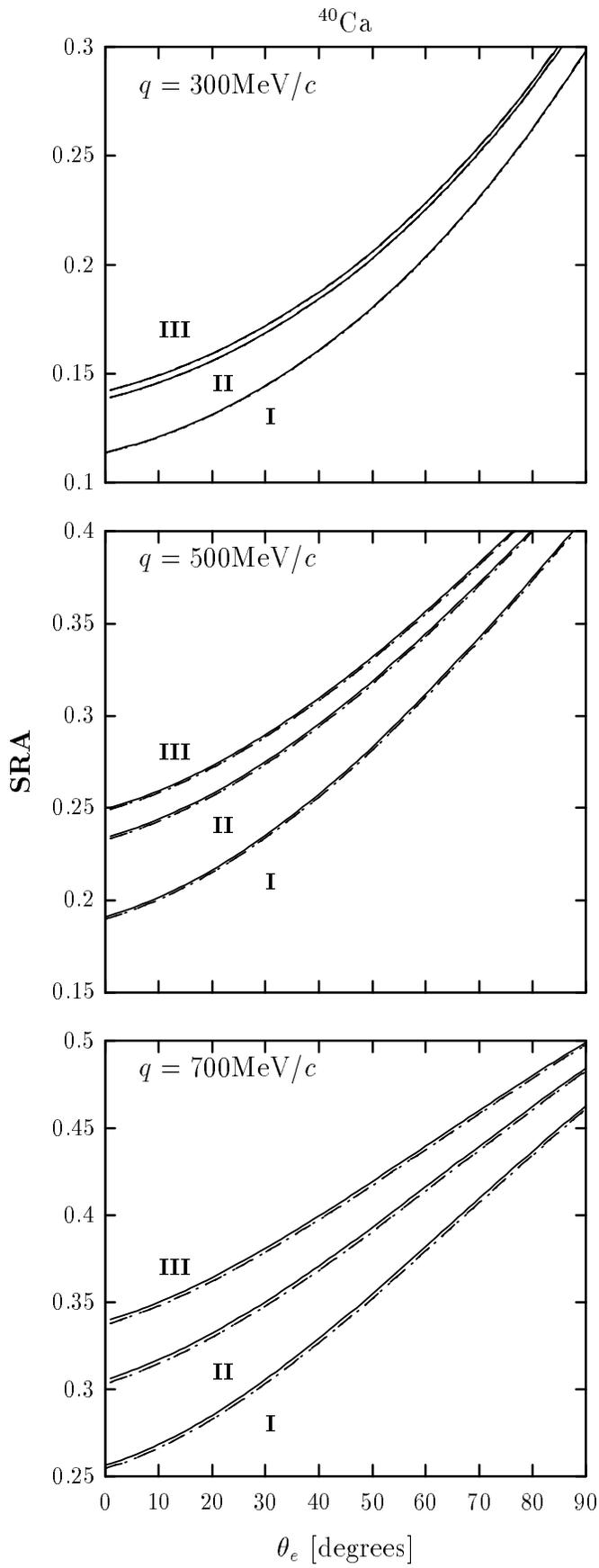
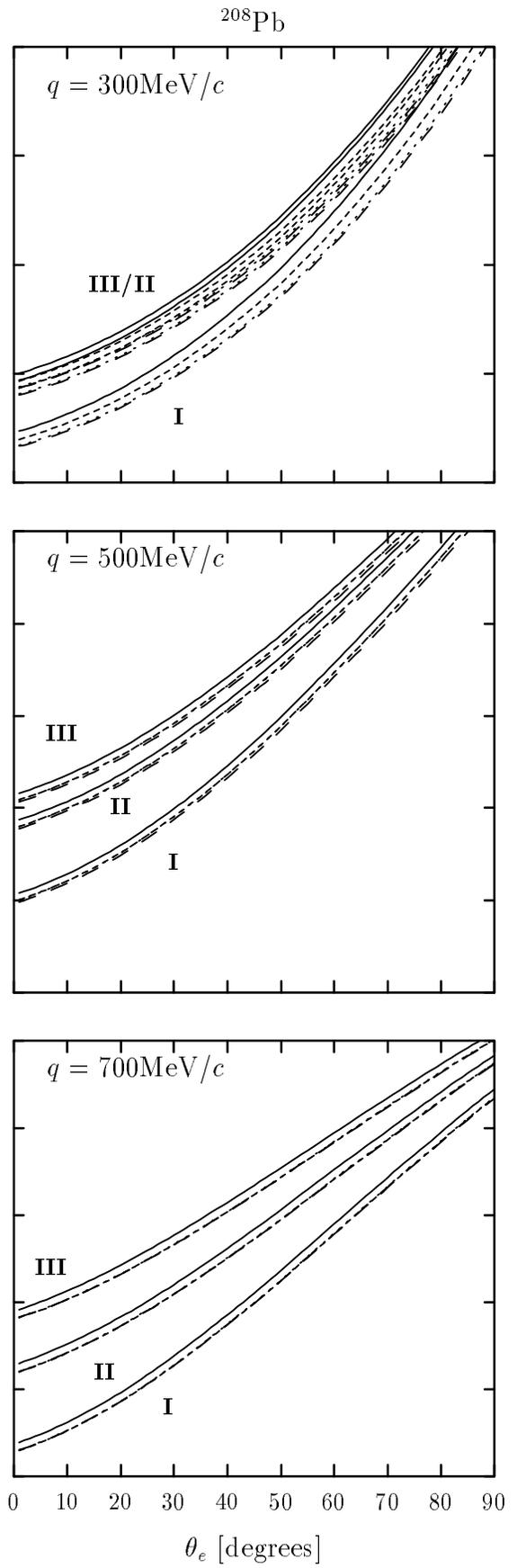
\end{document}